\documentclass{aa} % normal article
\usepackage{graphicx}
\usepackage{natbib}
\usepackage{longtable}
\usepackage[english]{babel}
\usepackage{txfonts}
\usepackage{color}

\begin{document}
   \title{MOJAVE: Monitoring of Jets in Active Galactic Nuclei with VLBA Experiments. IX. Nuclear opacity}% in compact AGN jets}
   \titlerunning{MOJAVE. IX. Nuclear opacity}
%   \subtitle{}

   \author{A.~B.~Pushkarev\inst{1,2,3}
          \and
          T.~Hovatta\inst{4,5}
          \and
          Y.~Y.~Kovalev\inst{6,1}
	  \and
	  M.~L.~Lister\inst{5}
          \and
          A.~P.~Lobanov\inst{1}
	  \and
	  T.~Savolainen\inst{1}
	  \and
	  J.~A.~Zensus\inst{1}
%	  \fnmsep
%	  \thanks{{\it Send offprint request to}: A. B. Pushkarev}
          }

   \institute{Max-Planck-Institut f\"ur Radioastronomie, Auf dem H\"ugel 69, 53121 Bonn, Germany\\
              \email{apushkar@mpifr.de}
         \and 
             Pulkovo Astronomical Observatory of the Russian Academy of Sciences, Pulkovskoe Chaussee 65/1, 196140 St. Petersburg, Russia
         \and
             Radio Astronomy Laboratory, Crimean Astrophysical Observatory, 98688 Nauchny, Crimea, Ukraine
	 \and
	     Cahill Center for Astronomy \& Astrophysics, California Institute of Technology, 1200 E. California Blvd, Pasadena, CA 91125, USA
	 \and
             Department of Physics, Purdue University, 525 Northwestern Avenue, West Lafayette, IN 47907, USA
         \and
             Astro Space Center of Lebedev Physical Institute of the Russian Academy of Sciences, Profsoyuznaya 84/32, 117997 Moscow, Russia 
%             \thanks{}
             }

   \date{Received 6 March 2012; accepted 20 July 2012}

% \abstract{}{}{}{}{} 
% 5 {} token are mandatory
 
  \abstract
  % context heading (optional); leave it empty if necessary  
   {}
   {We have investigated a frequency-dependent shift in the absolute position of the optically
     thick apparent origin of parsec-scale jets (``core shift'' effect) to probe physical
     conditions in ultra-compact relativistic outflows in active galactic nuclei.}
   {We used multi-frequency Very Long Baseline Array (VLBA) observations of 191 sources carried
     out in 12 epochs in 2006 within the Monitoring Of Jets in Active galactic nuclei with VLBA
     Experiments (MOJAVE) program. The observations were performed at 8.1, 8.4, 12.1, and
     15.4~GHz. We implemented a method of determining the core shift vector based on (i) image
     registration by two-dimensional normalized cross-correlation and (ii) model-fitting the
     source brightness distribution to take into account a non-zero core component offset from
     the phase center.}
   {The 15.4-8.1, 15.4-8.4, and 15.4-12.1~GHz core shift vectors are derived for 163 sources,
     and have median values of 128, 125, and 88~$\mu$as, respectively, compared to the typical
     measured errors of 50, 51, 35~$\mu$as. The effect occurs predominantly along the jet
     direction, with departures smaller than $45\degr$ from the median jet position angle in
     over 80\% of the cases. Despite the moderate ratio of the observed frequencies ($<$2), 
     core shifts significantly different from zero ($>$$2\sigma$) are detected for about 55\% 
     of the sources. These shifts are even better aligned with the jet direction, deviating from 
     the latter by less than $30\degr$ in over 90\% of the cases. There is an indication that the 
     core shift decreases with increasing redshift. Magnetic fields in the jet at a distance 
     of 1 parsec from the central black hole, calculated from the obtained core shifts, are found 
     to be systematically stronger in quasars (median $B_1\approx0.9$~G) than those in BL~Lacs (median
     $B_1\approx0.4$~G). We also constrained the absolute distance of the core from the apex of
     the jet at 15~GHz as well as the magnetic field strength in the 15~GHz core region.}
   {}

   \keywords{galaxies: active --
             galaxies: jets --
             quasars: general --
             radio continuum: galaxies
            }

   \maketitle
%
%________________________________________________________________

\section{Introduction}
\label{intro}

Bipolar relativistic outflows (jets) in active galactic nuclei (AGN)
are formed in the immediate vicinity of the supermassive central black
hole
%, at distances of $\sim$$10$ gravitational radii ($R_g=GM_{bh}/c^2$), 
and become detectable at distances of
$\gtrsim$$100$ gravitational radii ($R_g=GM_{bh}/c^2$) at millimeter 
wavelengths \citep{Junor_M87,Lobanov_07,Hada_M87}. The jets take away a
substantial fraction of the energy and angular momentum stored in the
accretion flow \citep{Hujeirat_03} and spinning central black hole
\citep{Koide_02,Komissarov_05}. As discussed by \cite{Vlahakis_04}, 
a poloidal-dominated magnetic field embedded in the accretion disk or in 
the black hole ergosphere is wound-up into toroidal loops that may 
provide effective jet collimation via hoop stress and accelerate the flow 
by magnetic pressure gradient up to a distance of $\sim$$10^3-10^5$~$R_g$.

Very Long Baseline Interferometry (VLBI) observations provide us with
the perfect zoom-in tool to explore AGN jets with a milliarcsecond
angular resolution corresponding to parsec-scale linear resolution. 
Typically, the parsec-scale radio morphology of a
bright AGN manifests a one-sided jet structure due to Doppler boosting
\citep[e.g.,][]{BlandfordKonigl79,Kellermann_07,MOJAVE} that enhances 
the emission of the approaching jet. The apparent base of the jet is 
commonly called the ``core'', and it is often the brightest and most 
compact feature in VLBI images of AGN. The VLBI core is thought 
to represent the jet region, located at the distance $r_\mathrm{core}$ 
to the central engine, at which its optical depth reaches $\tau_\nu\approx1$ 
at a given frequency. At short mm-wavelengths the core may also be the 
first recollimation shock downstream of the $\tau = 1$ surface instead 
of the surface itself. This does not affect our analysis, which uses 
longer wavelengths. Thus, the absolute position of the radio core is 
frequency-dependent and varies as $r_\mathrm{core}\propto\nu^{-1/k_r}$ 
\citep{BlandfordKonigl79,Koenigl81}, i.e., it shifts upstream at
higher frequencies and downstream at lower frequencies (the so-called
``core shift'' effect). The first core shift measurement from VLBI
observations was performed by \cite{Marcaide84}. Recent multi-frequency
studies of the core shift effect
\citep{Sullivan_09cs,Fromm10,Sokolovsky_11cs,Hada_M87} showed that
$k_r\approx1$ in most sources and epochs. This is consistent with the
\cite{BlandfordKonigl79} model of a synchrotron self-absorbed conical 
jet in equipartition between energy densities of the magnetic field and 
the radiating particle population. Nonetheless, departures in $k_r$ 
from unity are also possible and can be caused by pressure and density 
gradients in the jet or by external absorption from the surrounding 
medium \citep{L98,Kadler_04}.

The frequency-dependent offsets of the core positions can be used for
astrophysical studies of ultra-compact AGN jets to calculate the
magnetic fields, synchrotron luminosities, total (kinetic and magnetic
field) power, maximum brightness temperature and geometrical
properties of the jet \citep{L98}. The core shift effect also has
immediate astrometric applications. A typical shift between the radio
(4~cm) and optical (6000~\AA) domains for distant quasars is estimated
to be at the level of 0.1~mas \citep{Kovalev_cs_2008}, which is
comparable with the expected positional accuracy of the {\it GAIA}
astrometric mission \citep{Lindegren96}. Thus, the core shifts are
likely to influence not only the positional accuracy of the radio
reference frame but also an alignment of optical and radio astrometry
catalogs. Moreover, it is natural to expect that opacity properties
are variable on a time scale from months to years due to the
continuous emergence of new jet components, and especially during
strong nuclear flares. Therefore, as discussed by
\cite{Kovalev_cs_2008}, a special coordinated program is required to
perform multi-frequency and multi-epoch VLBI observation of a
pre-selected source sample to investigate the problem of core shift
variability.

A major difficulty in measuring the core shift is the accurate
registration of the VLBI images taken at different frequencies. The
problem stems from the loss of absolute position information in the
standard VLBI data reduction path, which involves self-calibration of
the station phases. Several approaches have been presented to
overcome this difficulty and measure core shifts. One of them is 
based on relative VLBI astrometry, i.e., phase-referencing to a
calibrator source \citep[e.g.][]{Marcaide84,Lara94,Guirado95,Ros01cs,
Bietenholz04,Hada_M87}. This particular technique is resource-consuming 
and has been used for a limited number of sources only. Another approach 
is the self-referencing method \citep{L98,Kovalev_cs_2008,Sokolovsky_11cs}, 
in which the core shift is derived by referencing the core position to 
bright optically thin jet features whose positions are expected to be 
achromatic. Although this method has provided the majority of known core 
shift measurements, it has a certain limitation. It cannot be applied 
for faint or smooth jets that lack compact bright feature(s) well
separated from the core at different frequencies. A proper alignment
of the optically thin parts of the jet can also be accomplished by 
two-dimensional cross-correlation of the images, initially suggested 
and performed by \cite{Walker_2D} for multi-frequency VLBA observations 
of 3C~84.  The algorithm was also discussed by \cite{Croke_2D} and Fromm 
et al. (in prep.), and applied by \cite{Sullivan_09cs} to obtain core 
shifts in four BL~Lac objects. This approach, in conjunction with source 
model fitting, presents a more widely applicable method for deriving core
shifts (see Sect.~\ref{s:method} for detailed discussion), which we
use in this paper. Another alternative indirect method recently proposed 
by \cite{Kudryavtseva_11_cs} is based on an analysis of time lags of 
flares monitored with single-dish observations. Although it has obvious
limitations on the epoch at which the core shift can be measured, the
method is promising for highly compact sources, which pose problems for 
other opacity study methods due to the lack of optically thin jet structure. 
It is noteworthy that all of the aforementioned techniques provide a 
comparable accuracy level. To date, only two core shift studies 
\citep{Kovalev_cs_2008,Sokolovsky_11cs} have been carried out on large 
samples. They have shown that the effect is significant for many sources.

In this paper, we measure frequency-dependent shifts in the absolute 
core positions and study the statistical properties of the detected
core shift vectors by using a large sample of sources from the MOJAVE 
(Monitoring Of Jets in Active galactic nuclei with VLBA Experiments) 
program \citep{MOJAVE}. We also analyze systematics and discuss the 
uncertainties of the two-dimensional cross-correlation technique, investigating its 
properties for different jet morphologies. We constrain the basic physical 
properties of the jets, such as the magnetic field strength in the core 
region and at the true base of the flow, the distance from the jet apex 
to the radio core, as well as the estimate of the central black hole mass 
from the derived core shifts.

Throughout the paper, we assume the power index $k_r=1$, i.e.,
$r_\mathrm{core}\propto\nu^{-1}$ (see model assumptions for this case above). 
We use the $\Lambda$CDM cosmological model with $H_0=71$~km~s$^{-1}$~Mpc$^{-1}$,
$\Omega_m=0.27$, and $\Omega_\Lambda=0.73$ \citep{Komatsu09}. All
position angles are given in degrees from north through east.

\section{Observations and data processing}

The MOJAVE project \citep{MOJAVE} is a long-term VLBA program aimed at
investigating the structure and evolution of extragalactic
relativistic radio jets in the northern sky. The full monitoring list
currently consists of about 300 sources, and includes a statistically
complete, flux-density limited sample of 135 AGN, referred to as
MOJAVE-1. In addition to the program's usual single-frequency setup at
15.4~GHz, 12 monthly separated epochs of observations during 2006 were
carried out simultaneously also at 12.1, 8.4, and 8.1~GHz. The
observations were made in dual circular polarization mode, with a
bandwidth of 16~MHz at two lower bands and 32~MHz at two upper bands,
and recorded with a bit rate of 128~Mbit~s$^{-1}$. In total, 191
sources were observed.

The initial calibration was performed with the NRAO Astronomical Image
Processing System (AIPS) \citep{aips} following the standard techniques. 
All frequency bands were processed separately throughout the data reduction. 
CLEANing \citep{CLEANref}, phase and amplitude self-calibration 
\citep{Jennison58,Twiss_etal60}, were performed in the Caltech Difmap 
\citep{difmap} package. In all cases a point-source model was used as 
an initial model for the iterative procedure. Final maps were produced 
by applying natural weighting of the visibility function. For a more 
detailed  discussion of the data reduction and imaging process schemes, 
see \cite{MOJAVE,Hovatta_RM_11}.

The structure of each source at each frequency band was model-fitted
in the visibility ($u,\upsilon$) plane in Difmap using circular and
elliptical Gaussian components. To achieve matched resolution in all 
bands in the image plane, we appropriately cut the long baselines 
from the 15.4 and 12.1~GHz interferometric visibility data sets and short
baselines from 8.1 and 8.4~GHz data sets. For each source, all maps
were restored with the same beam size taken from lowest frequency
(8.1~GHz) data using a pixel size of 0.03~mas, and these images were 
cross-correlated to register them as explained in Sect.~\ref{s:method}.

This set of observations was also used to investigate the jet Faraday
rotation measures \citep{Hovatta_RM_11} and spectral index
distributions (Hovatta et al., in prep.).

\section{Method for measuring the core shift}
\label{s:method}

\begin{figure}
\begin{center}
 \resizebox{0.95\hsize}{!}{\includegraphics[angle=-90,clip=true]{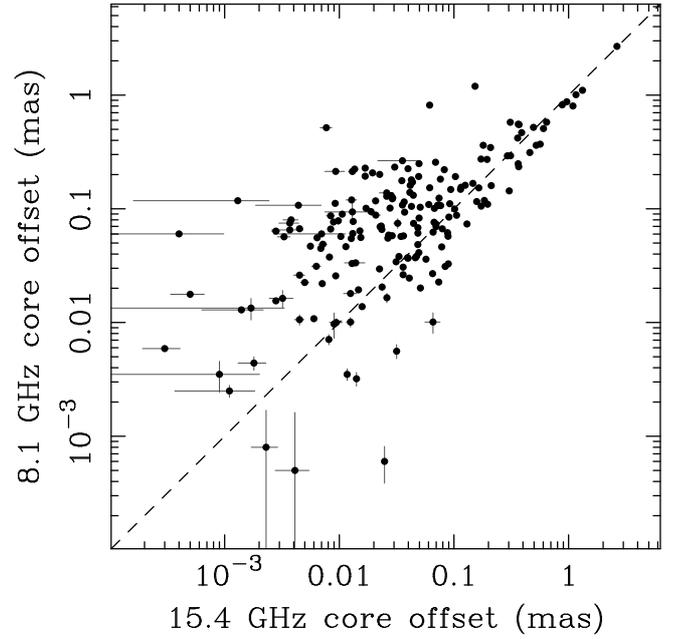}}
\end{center}
 \caption{ Phase center offsets of the core components at 8.1 GHz
   versus that at 15.4 GHz taken from model fits.  }
 \label{f:core_offsets}
\end{figure}

Registration of two VLBI images taken at different frequencies
provides a shift between the image phase centers, such that they are
co-aligned to the same position on the sky. This image shift is not
equal to the core shift we want to measure. 
%
%It would be equal only if
%the core component relative position in the image plane does not
%change with frequency, for instance it strickly coincides with the
%phase center. In turn, the latter is possible only for a completely
%unresolved source, because the absolute position information is lost
%during the phase self-calibration procedure, which places the peak of
%intensity at the origin of coordinates.  
%
In general, the presence of
the jet structure we detect results in a non-zero offset of the core
from the phase center. As the core components significantly dominate
in total flux, the magnitude of the offsets is typically small, but at
the same time not negligible. When a distant jet component is brighter
than the core, the offset can be large, as in the case of the quasar
0923+392, where the offset is about 2.6~mas. Extreme cases are
discussed in \cite{Petrov_11}. Moreover, these offsets are different
at different frequencies for a given source, and become statistically
larger at lower frequencies due to spectral properties of the
jets. Figure~\ref{f:core_offsets} shows the core offsets at 15.4 and
8.1~GHz, with the median values being 36 and 81~$\mu$as, respectively.

The core position departures from the phase center thus have to be 
taken into account to derive the core shifts. Because the image shift
\vec{IS} is independent of the core component position, it measures
the vector sum of the absolute core shift \vec{CS} ($|\vec{CS}|=\Delta
r_\mathrm{core,\,\nu_1\nu_2}$) and the difference in the coordinates
(offset shift relative to the map center) \vec{OS}:
\begin{equation}
\vec{IS} = \vec{CS} + \vec{OS}\,,
\label{e:vector_eq}
\end{equation}
from which the magnitude and direction of the core shift vector can be
readily calculated.

We used the fast normalized cross-correlation (NCC) algorithm by
\citet{2D_cross_corr} to register the images across the
frequencies. The algorithm allows one to apply frequency domain
methods to calculate the unnormalized cross correlation and then
efficiently normalizes it by using precomputed integrals of the
images over the search area. Thus, spatial domain computation of the
cross correlation function is not needed and for large images the
decrease in computing time is significant. The features that were
matched between the images were selected from the optically thin
part of the jet and assumed to have constant spectral index across
them. The effect of possible spectral index gradients across the
features is discussed in Sect.~\ref{s:errors}. In all cases, the
image shifts $\vec{IS}$, obtained with NCC, were verified by visually
inspecting the corresponding spectral index images before and after
the alignment, because the latter are extremely sensitive to the
(in)accurate image alignment, as was shown by \cite{Kovalev_cs_2008}. 
The spectral index images are presented and discussed by Hovatta et 
al. (in prep.). The image shifts between 15.4~GHz and other bands are 
found to be within a range of 0 and 1.11~mas, with a median value of 
0.13~mas. The extreme value of 1.11~mas was detected in 3C~273 between 
15.4 and 8.4~GHz, where the peak of brightness at those two frequencies 
corresponds to different features in the source structure. In some rare 
cases (e.g., 1928+738 at epoch 2006 Apr 28 of and 2128$-$123 at epoch of 
2006 Oct 6), when an optically thin, bright, compact jet component 
dominates in flux density, the image shift was measured to be zero, as 
expected for achromatic jet components.  The accuracy of the two-dimensional 
cross-correlation technique is discussed in Sect.~\ref{s:errors}.

The magnitude of the core offset difference vector $\vec{OS}$ between
15.4~GHz and other bands ranged between 0 and 1.09~mas, with a median
value of 0.05~mas. The maximum value of 1.09~mas holds also for quasar 
3C~273 due to the same reason.

\begin{table*}
\caption{Derived core shift vectors.}
\label{t:core_shift}
\centering
\begin{tabular}{ccrrrrrrrrrr}
\hline\hline

  Source   &      Epoch &   Median & \multicolumn{3}{c}{15.4-8.1 GHz core shift} & \multicolumn{3}{c}{15.4-8.4 GHz core shift} & \multicolumn{3}{c}{15.4-12.1 GHz core shift} \\
           &            &   jet PA &    PA  &    total &      proj &      PA  &  total &      proj &      PA  &  total &      proj \\
           &            &    (deg) &  (deg) &    (mas) &     (mas) &    (deg) &  (mas) &     (mas) &    (deg) &  (mas) &     (mas) \\
   (1)     &        (2) &      (3) &    (4) &      (5) &       (6) &      (7) &    (8) &       (9) &     (10) &   (11) &      (12) \\
\hline
0003$-$066 & 2006-07-07 &  $-$82.1 &  $-$60.3 &  0.035 &     0.033 &  $-$25.8 &  0.019 &     0.011 &  $-$62.5 &  0.015 &     0.014 \\
0003$+$380 & 2006-03-09 &    117.0 &     77.2 &  0.134 &     0.103 &     79.6 &  0.139 &     0.110 &     77.2 &  0.124 &     0.095 \\
0003$+$380 & 2006-12-01 &    116.2 &    115.5 &  0.063 &     0.063 &    121.7 &  0.106 &     0.106 &    103.0 &  0.046 &     0.044 \\
0007$+$106 & 2006-06-15 &  $-$67.6 &  $-$88.5 &  0.008 &     0.007 &  $-$20.1 &  0.011 &     0.007 &    146.2 &  0.008 &  $-$0.007 \\
0010$+$405 & 2006-04-05 &  $-$31.8 &  $-$38.3 &  0.013 &     0.013 &     64.7 &  0.008 &  $-$0.001 &   \ldots & \ldots &    \ldots \\
0010$+$405 & 2006-12-01 &  $-$32.6 &   $-$0.3 &  0.005 &     0.004 &     39.1 &  0.005 &     0.001 &  $-$89.9 &  0.010 &     0.006 \\
0055$+$300 & 2006-02-12 &  $-$50.1 &  $-$45.2 &  0.179 &     0.179 &  $-$10.2 &  0.083 &     0.064 &  $-$61.5 &  0.053 &     0.052 \\
0106$+$013 & 2006-07-07 & $-$125.2 & $-$113.7 &  0.005 &     0.005 & $-$139.2 &  0.005 &     0.005 &     20.2 &  0.002 &  $-$0.001 \\
%0109$+$224 & 2006-05-24 &     85.1 &     65.3 &  0.147 &     0.138 &     91.6 &  0.073 &     0.072 &     75.8 &  0.120 &     0.118 \\
%0111$+$021 & 2006-03-09 &    129.8 &    114.4 &  0.137 &     0.132 &    158.1 &  0.174 &     0.154 &    140.4 &  0.087 &     0.086 \\
\hline
\end{tabular}
\tablefoot{
%Columns are as follows:
(1) IAU name (B1950.0);
(2) epoch of observations;
(3) 15.4~GHz median jet position angle;
(4) position angle of the 15.4-8.1~GHz core shift vector;
(5) magnitude of the 15.4-8.1~GHz core shift vector;
(6) 15.4-8.1~GHz core shift vector in projection on the median position angle;
(7), (8), (9), and (10), (11), (12) the same as (4), (5), (6) but for 15.4-8.4~GHz and 15.4-12.1~GHz 
core shifts, respectively.
Table~\ref{t:core_shift} is published in its entirety in the electronic version of {\it Astronomy 
\& Astrophysics}. A portion is shown here for guidance regarding its form and content.
}
\end{table*}

\begin{figure}
\begin{center}
 \resizebox{0.88\hsize}{!}{\includegraphics[angle=-90,clip=true]{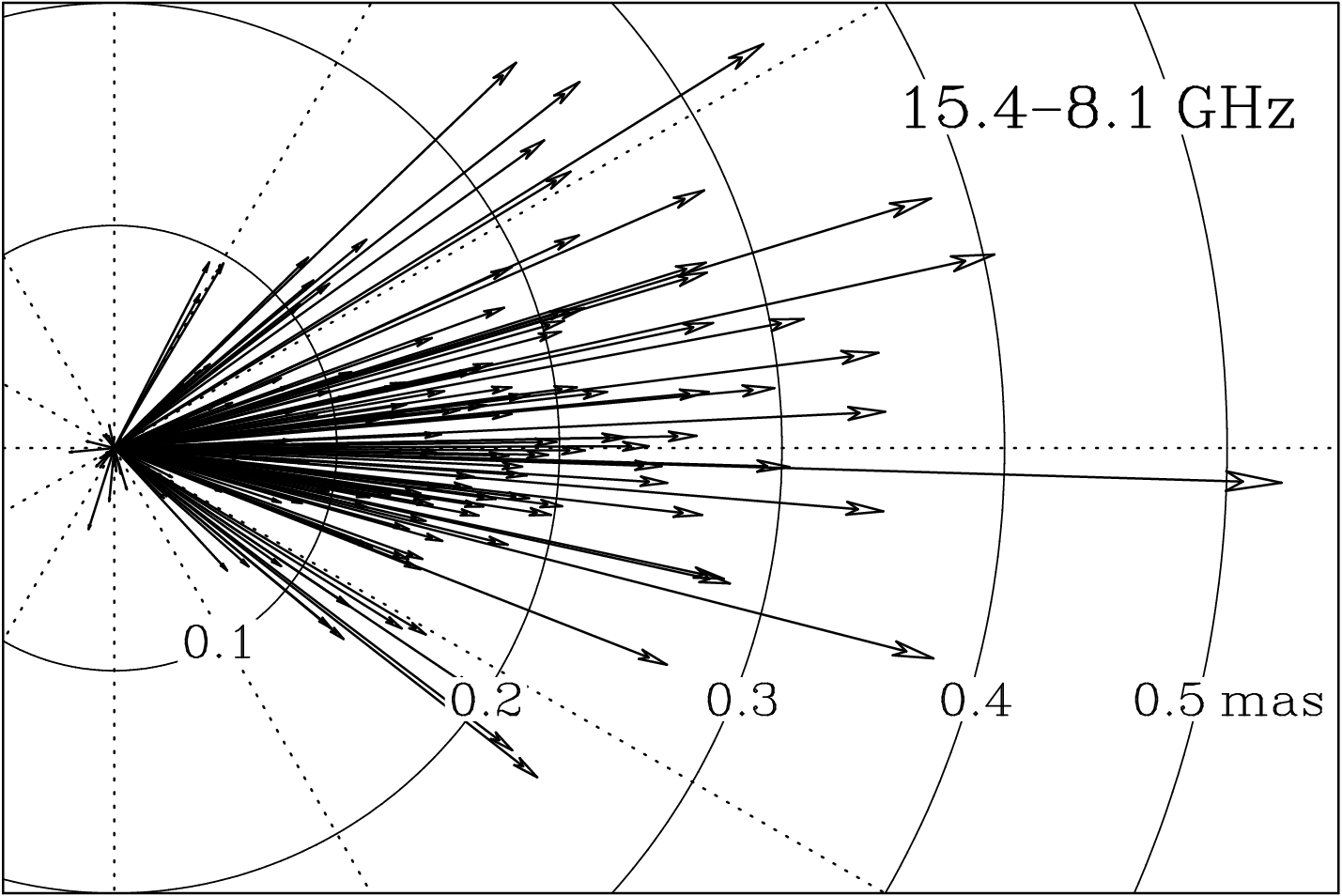}}\vspace{1mm}
 \resizebox{0.88\hsize}{!}{\includegraphics[angle=-90,clip=true]{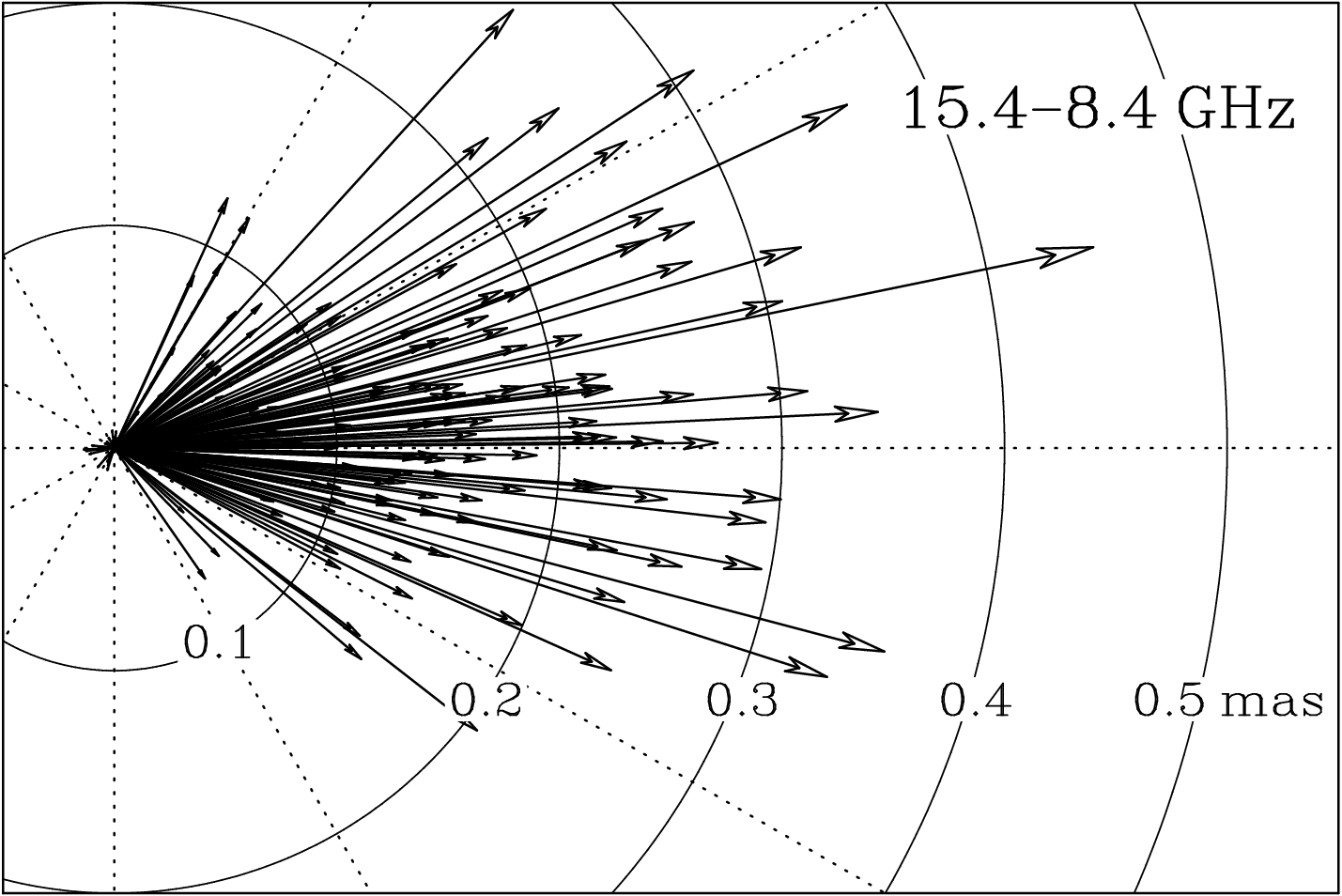}}\vspace{1mm}
 \resizebox{0.88\hsize}{!}{\includegraphics[angle=-90,clip=true]{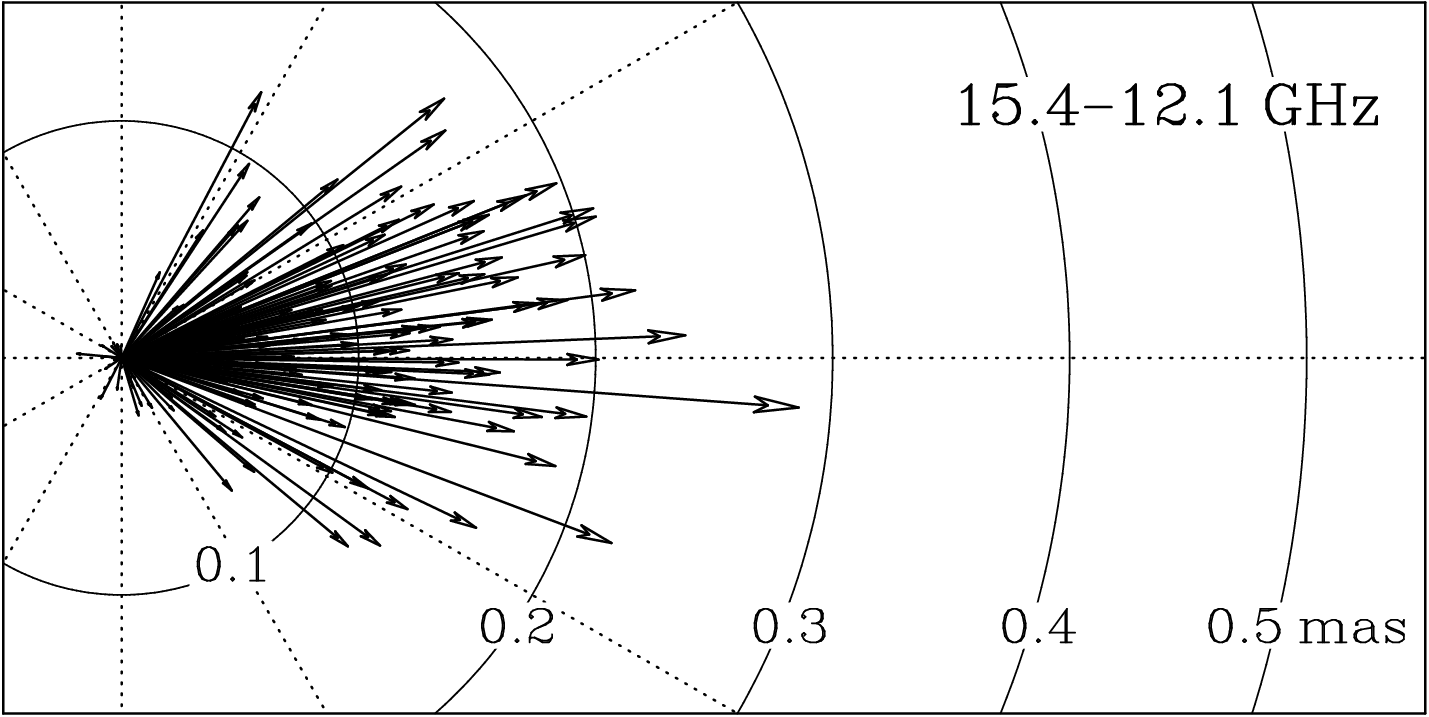}}
\end{center}
 \caption{ Polar plots of the 15.4-8.1 (top), 15.4-8.4 (middle), and
   15.4-12.1~GHz (bottom) core shift vectors.  The polar axis is
   pointing to the right and co-aligns with the median jet direction
   of a source.  Radial distance is given in mas. Dotted lines are
   drawn at intervals of $30\degr$.  }
 \label{f:core_shift_polar}
\end{figure}

\section{Core shift measurement results}
\label{s:results}

Substituting the results of model fitting and two-dimentional cross correlation
into Eq.~(\ref{e:vector_eq}), we calculated the magnitude $\Delta
r_\mathrm{core,\,\nu_1\nu_2}$ and direction $\theta_\mathrm{cs}$ of
the 15.4-8.1, 15.4-8.4, and 15.4-12.1~GHz core shift vectors for 160,
158, and 147 sources (Table~\ref{t:core_shift}), respectively. For 
the other 31, 33, and 44 sources we could not measure the respective
15.4-8.1, 15.4-8.4, and 15.4-12.1 core shifts, mostly due to the
weakness of their jet emission (especially at 15.4~GHz). This made the 
cross-correlation technique inapplicable, since there was no sufficiently
large optically thin emission structure for feature matching. We also 
excluded those sources, mostly nearby galaxies listed in 
Table~\ref{t:excluded_sources}, whose core region was complex (e.g., 
3C~84, M~87; the core shift in M~87 was studied by \citealt{Hada_M87} 
using phase-referencing VLBA observations) or for which the identification 
of the core component was unclear (e.g., 0108+388, 1509+054). The maximum 
and median magnitude of the derived 15.4-8.1, 15.4-8.4, and 
15.4-12.1~GHz core shift vectors in angular and linear scale are 
summarized in Table~\ref{t:core_shift_stat}. As seen from
Table~\ref{t:core_shift_stat}, the median values of the 15.4-8.1 and
15.4-8.4~GHz core shifts are comparable, while the 15.4-12.1~GHz ones 
are statistically smaller, as expected. In angular scale, these values 
are of about 8\% of the corresponding FWHM beam size at 8.1~GHz.

In Fig.~\ref{f:core_shift_polar}, we present plots of the derived core
shifts in polar coordinates, where the head of each core shift vector
represents the core position at lower frequency, while all core
positions at higher frequency are placed at the origin. The polar axis
corresponds to the median jet direction $\theta_\mathrm{jet}$
calculated from position angles of the jet components with respect to
the core component, using a corresponding model fit at 15.4~GHz. Thus
the polar coordinates of the head of each vector represent the
magnitude of the core shift vector, $\Delta
r_\mathrm{core,\nu_1\nu_2}$, and the angular deviation from the jet
direction, $\theta_\mathrm{cs}-\theta_\mathrm{jet}$. The shift effect
occurs predominantly along the jet direction. In more than 80\% of cases,
the core shift vectors deviate less than $45\degr$ from the median jet
position angle. Statistically, the larger core shifts have better
alignment with the jet direction because (i) they are less influenced
by random errors and (ii) the core shift takes place along the jet in
most cases.  The weighted average of
$\theta_\mathrm{cs}-\theta_\mathrm{jet}$ is close to zero. Significant
angular deviations of the core shift vectors from the median jet
direction may take place in sources with substantial jet bending,
either within an unresolved region near the VLBI core, or in
the outer jet, thus affecting the median jet position angle. We have
also analyzed distributions of the angular deviation between the core
shift vectors and (i) inner jet direction determined as a position
angle of the innermost jet component at 15.4~GHz and (ii) flux
density-weighted average of the position angles of all Gaussian jet
components at 15.4~GHz. In both these cases the scatter was larger,
indicating that the median jet position angle is a better estimate 
for the direction of the outflow for the majority of sources.

\begin{table}
\caption{Sources excluded from the analysis due to unclear core position.}
\label{t:excluded_sources}
\centering
\begin{tabular}{c l c l}
\hline\hline
Source     & Alias    &Opt. class$^a$ & ~~~$z^a$   \\
\hline
0108$+$388 &          &    G    & 0.668    \\
0238$-$084 & NGC 1052 &    G    & 0.005037 \\
0316$+$413 & 3C 84    &    G    & 0.0176   \\
0429$+$415 & 3C 119   &    Q    & 1.022    \\
0710$+$439 &          &    G    & 0.518    \\
1228$+$126 & M 87     &    G    & 0.00436  \\
1404$+$286 & OQ 208   &    G    & 0.077    \\
1509$+$054 &          &    G    & 0.084    \\
1607$+$268 & CTD 93   &    G    & 0.473    \\
1957$+$405 & Cygnus A &    G    & 0.0561   \\
2021$+$614 & OW 637   &    G    & 0.227    \\
\hline
\end{tabular}
\tablefoot{$^a$ As compiled by \cite{MOJAVE}}
\end{table}

\begin{table}
\caption{Core shift statistics.}
\label{t:core_shift_stat}
\centering
\begin{tabular}{c c c c c c c c c}
\hline\hline
Core shift    && \multicolumn{2}{c}{angular} && \multicolumn{2}{c}{linear} \\
              &  N  &       max &          med &  N  &   max &   med \\
              &     & ($\mu$as) &    ($\mu$as) &     &  (pc) &  (pc) \\
\hline
15.4-8.1~GHz  & 160 &       525 &          128 & 154 & 3.108 & 0.658 \\
15.4-8.4~GHz  & 158 &       449 &          125 & 152 & 2.927 & 0.643 \\
15.4-12.1~GHz & 147 &       286 &\phantom{1}88 & 142 & 2.012 & 0.429 \\
\hline
\end{tabular}
\end{table}

Our median 15.4-8.1~GHz core shift of 0.128~mas exceeds that of
0.080~mas reported by \cite{Sokolovsky_11cs}, which was based on a 
smaller sample of 20 sources, for which the core shifts were derived 
using the self-referencing method.  But if the 15.4-8.1~GHz core shifts 
for the 20-source sample are calculated from the fitted hyperbolas
\citep{Sokolovsky_11cs}, which provide more accurate core shift values,
the corresponding median yields 0.127~mas, which agrees well with the
median of our sample.

\begin{figure}
\begin{center}
\resizebox{0.94\hsize}{!}{\includegraphics[angle=-90,clip=true]{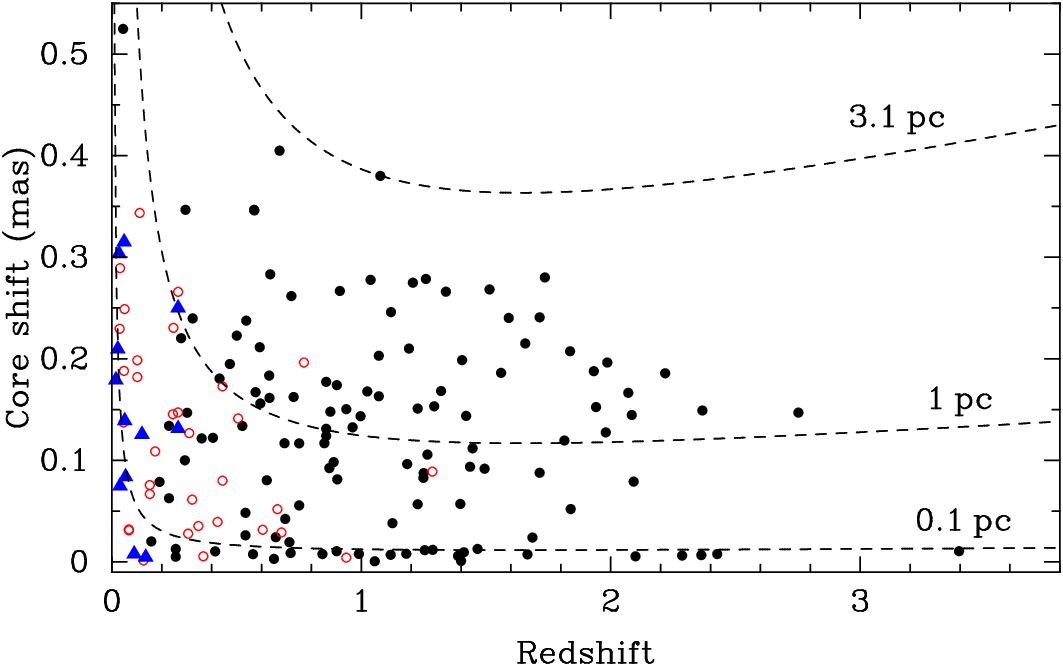}}\vspace{1mm}
\resizebox{0.94\hsize}{!}{\includegraphics[angle=-90,clip=true]{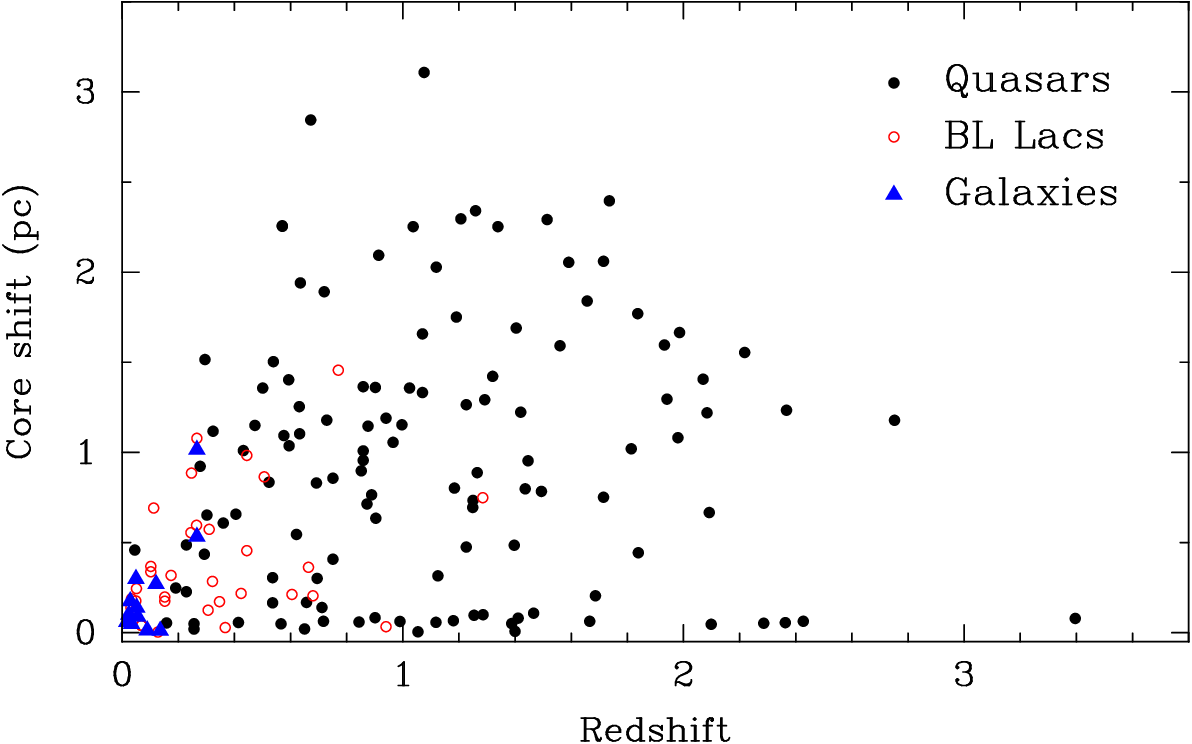}}
\end{center}
 \caption{ Top: 15.4-8.1~GHz core shift in angular scale versus
   redshift. The dashed lines correspond to a fixed projected linear scale of
   0.1, 1, and 3.1~pc. The latter curve envelopes all the measurements.
   Bottom: 15.4-8.1~GHz core shifts in projected linear scale versus
   redshift.  Typical error values are discussed in
   Sect.~\ref{s:errors}. }
 \label{f:core_shift_vs_z}
\end{figure}

The sources with the largest angular shifts are all at $z<1$, as shown
in Fig.~\ref{f:core_shift_vs_z} (top), where we plot the 15.4-8.1~GHz
core shift against redshift values. All measurements are confined
under the aspect line corresponding to 3.1~pc, the maximum linear
15.4-8.1~GHz shift (see Table~\ref{t:core_shift_stat}). In contrast,
in linear projected scale (Fig.~\ref{f:core_shift_vs_z}, bottom), 
low-redshift sources are characterized by small shifts due to quick 
falling of the scale factor, while more distant sources have larger 
shifts. We have found evidence that the angular core shift decreases
with increasing redshift by binning the data into nine equally 
populated bins, though we cannot claim the dependence to be highly 
significant due to the large scatter in measured core shifts, which 
stems from different physical conditions and viewing angles in 
different sources at the same redshift. The plots for 15.4-8.4 and
15.4-12.1~GHz core shifts are qualitatively similar to 
Fig.~\ref{f:core_shift_vs_z}.

Additionally, we tested whether the core shift measurements are affected 
by limited angular resolution blending following the approach used by 
\cite{Kovalev_cs_2008}. If present, this effect would preferentially
increase the magnitude of the core shift vectors when they are better
aligned with the major axis of the interferometric restoring beam resulting 
in a U-shape dependence between $\Delta r_\mathrm{core,\,\nu_1\nu_2}$ and
$|\theta_\mathrm{cs}-\theta_{\mathrm{bpa}}|$, where $\theta_{\mathrm{bpa}}$ 
is the position angle of the major axis of the beam, and 
$-90\le\theta_{\mathrm{bpa}}\le90$. We found no such trend in the 
15.4-8.1~GHz measurements, confirming that the registered core shifts 
are not dominated by blending.

\section{Accuracy of the method}
\label{s:errors}
\subsection{Random errors}

\begin{figure}
\begin{center}
\resizebox{0.97\hsize}{!}{\includegraphics[angle=-90,clip=true]{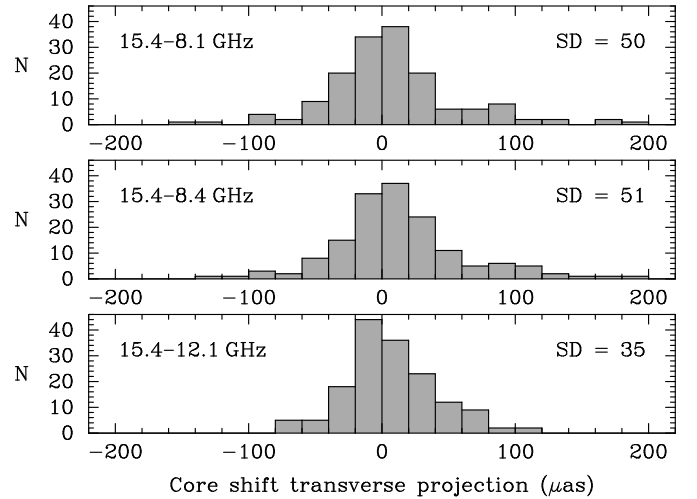}}
\end{center}
 \caption{ Distributions of transverse projections of the 15.4-8.1,
   15.4-8.4, and 15.4-12.1~GHz core shift vectors onto the jet
   direction.  }
 \label{f:cs_err}
\end{figure}

As seen from Eq.~(\ref{e:vector_eq}), the uncertainty of the core
shift $\sigma_{\nu_1\nu_2}$ is determined by errors of the core
positions and image registration. However, individual {\it a
posteriori} estimates of the image registration accuracy are
problematic, since (1) in our case the values of the applied
similarity criterion (NCC) are not normally distributed, making
an estimation of random errors difficult, and since (2) systematic
errors can only be addressed by simulations. Therefore, we used a
statistical approach to assess the typical random core shift
error in our sample. If we assume that (i) the core shift
occurs along the jet and (ii) the errors are random in direction and
comparable to each other, then the standard deviation of the
transverse projections of the core shift vectors onto the jet
direction yields the typical error. In Fig.~\ref{f:cs_err}, we plot
the corresponding 15.4-8.1, 15.4-8.4, and 15.4-12.1~GHz distributions,
from which we find $\sigma_\mathrm{15.4-8.1\,GHz}=50$~$\mu$as,
$\sigma_\mathrm{15.4-8.4\,GHz}=51$~$\mu$as, and
$\sigma_\mathrm{15.4-12.1\,GHz}=35$~$\mu$as. From these values, we
determine that in 57\%, 59\%, and 58\% of cases the respective core
shifts are significantly ($>$$2\sigma$) different from zero, more than 
90\% of which in turn deviate less than $30\degr$ from the median jet 
direction. The derived error estimates are conservative, since in some 
cases the angular deviations of the core shift vector from the median 
jet direction can be real, for instance, in curved jets.

An alternative way to estimate a typical random error in core
shift is based on the fact that 8.1 and 8.4~GHz bands are closely
separated, but were processed independently. Therefore, the 15.4-8.1
and 15.4-8.4~GHz core shifts are expected to be virtually the same and
the non-zero difference between them is due purely to errors. This
approach was also used in \cite{Sokolovsky_11cs}. Constructing a
distribution of these differences and calculating its standard
deviation, we found
$\sigma_\mathrm{15.4-8.1\,GHz}\approx\sigma_\mathrm{15.4-8.4\,GHz}\approx55$~$\mu$as,
which is consistent with the error estimates obtained using the first
method.

Since there is some freedom in selecting the jet feature to be
matched in NCC, there is a possibility that user decisions affect 
the image registration results. We tested the robustness of the
registration algorithm to ``user bias'' by having two different
people separately perform the image alignment for one of the
observing epochs. They both had similar instructions regarding the
selection of the matched feature, i.e., (1) it should be optically
thin, (2) and it should have as much structural variation as
possible. The distribution of the differences in \vec{IS} had a
standard deviation of $\approx 50$~$\mu$as for $N=24$ pairs of
images. This also closely matches the typical random error estimated
above.

\begin{figure}
\begin{center}
\resizebox{0.66\hsize}{!}{\includegraphics[angle=-90,clip=true]{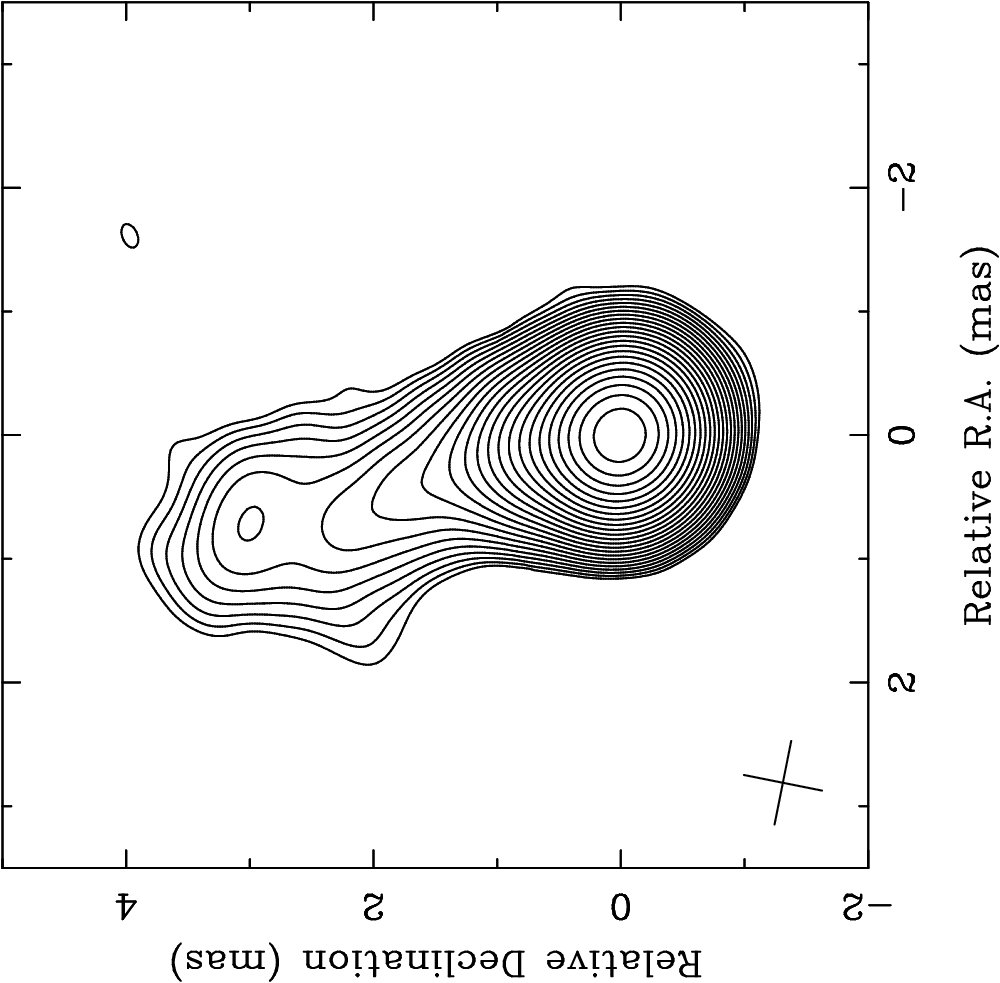}}
\resizebox{0.70\hsize}{!}{\includegraphics[angle=-90,clip=true]{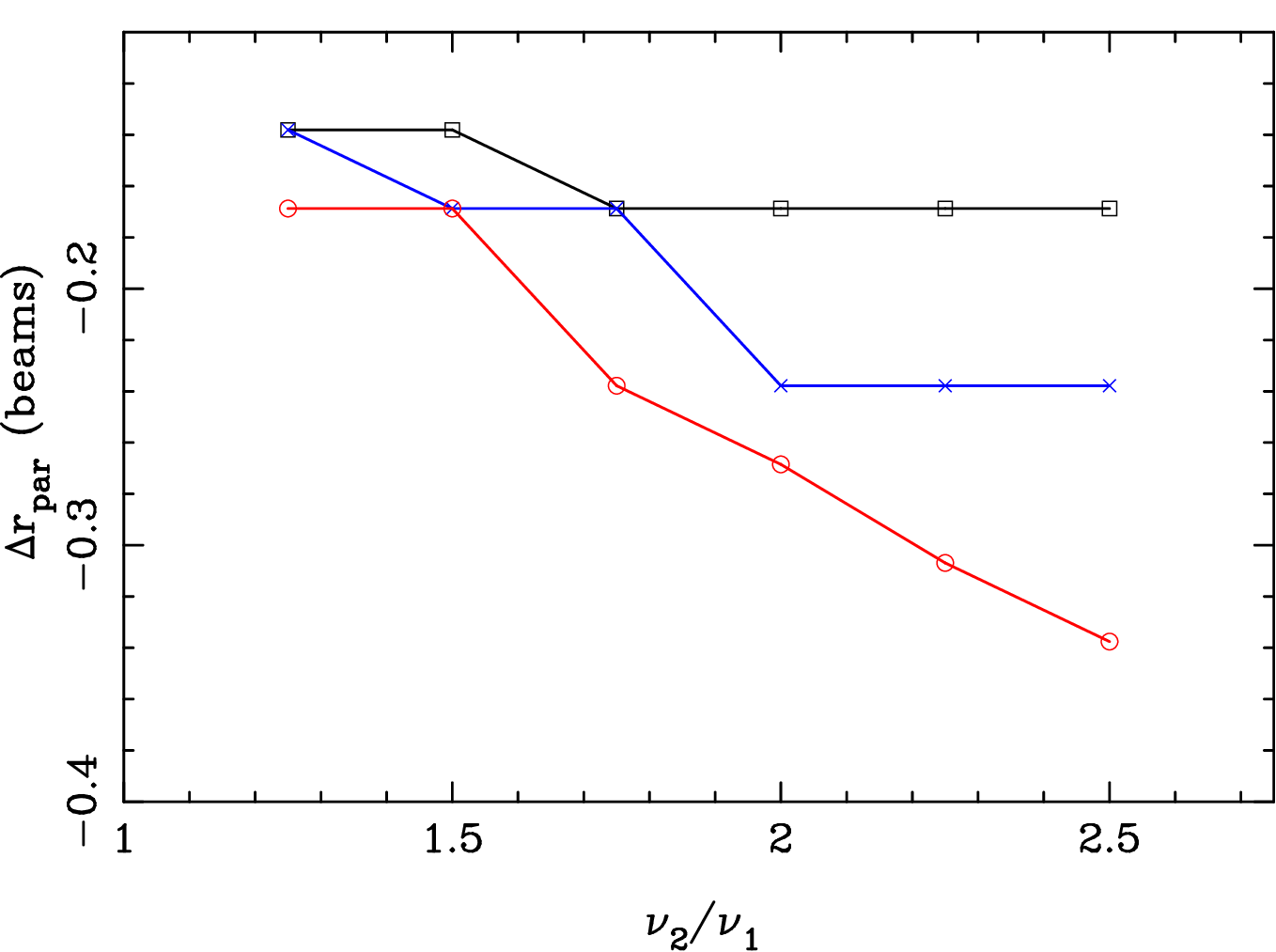}}
\resizebox{0.70\hsize}{!}{\includegraphics[angle=-90,clip=true]{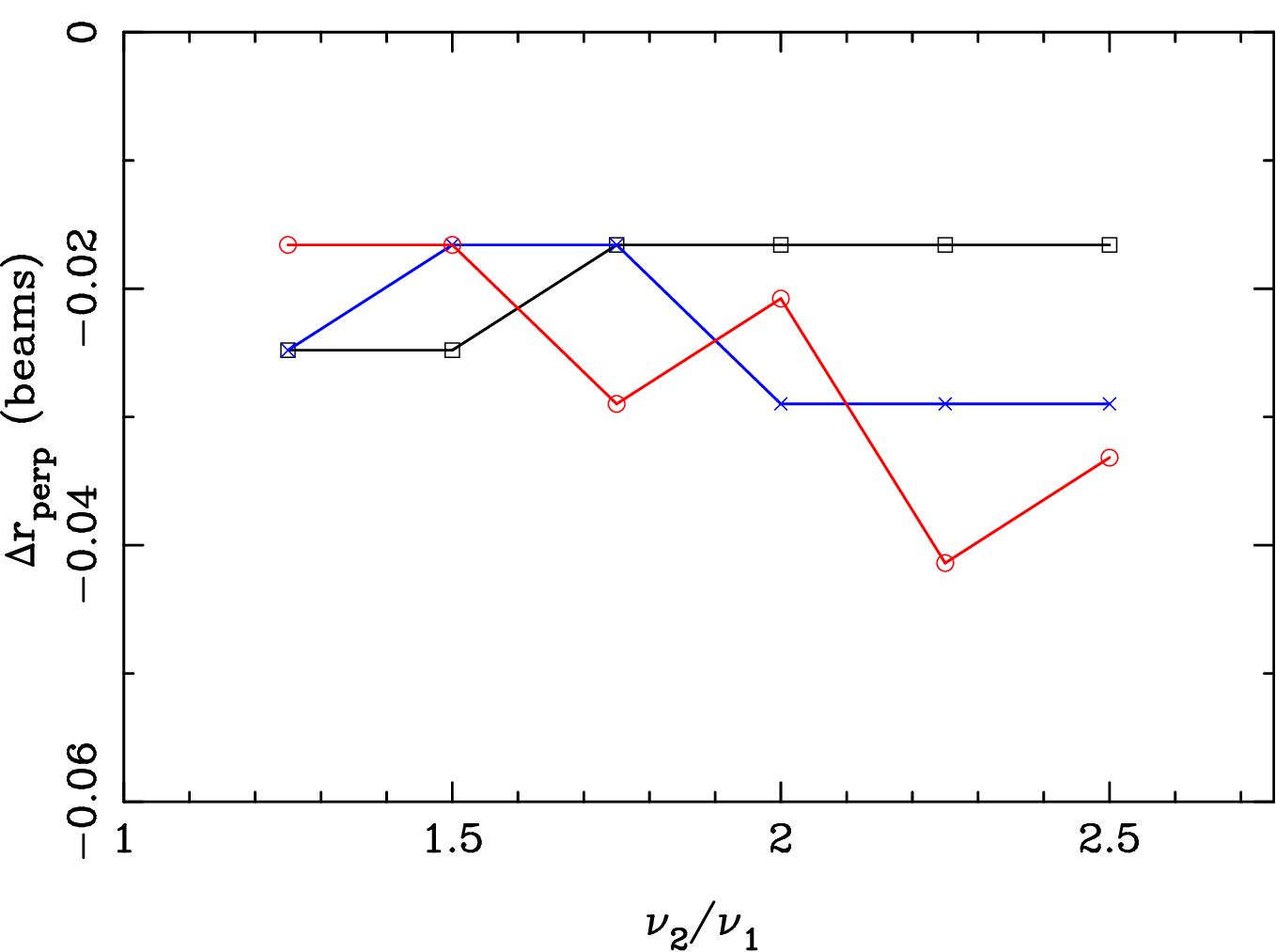}}
\end{center}
 \caption{ Naturally weighted total intensity CLEAN image of 0716+714 at 15.4~GHz (top). 
           The alignment error (in beam widths) along the jet as a function of 
           frequency ratio (middle). The alignment error (in beam widths) perpendicular 
           to the jet (bottom). The error is defined as $\vec{r_{\nu_1}}-\vec{r_{\nu_2}}$ 
           where $\vec{r_{\nu_1}}$ is the position in the low-frequency image and 
           $\vec{r_{\nu_2}}$ is the corresponding position in the high-frequency image. 
           The different spectral index gradients are shown in different colors: 
           $-$0.1~mas$^{-1}$ (black squares), $-$0.2~mas$^{-1}$ (blue crosses), and $-$0.3~mas$^{-1}$ (red circles). }
 \label{f:0716+714}
\end{figure}

\begin{figure}
\begin{center}
\resizebox{0.65\hsize}{!}{\includegraphics[angle=-90,clip=true]{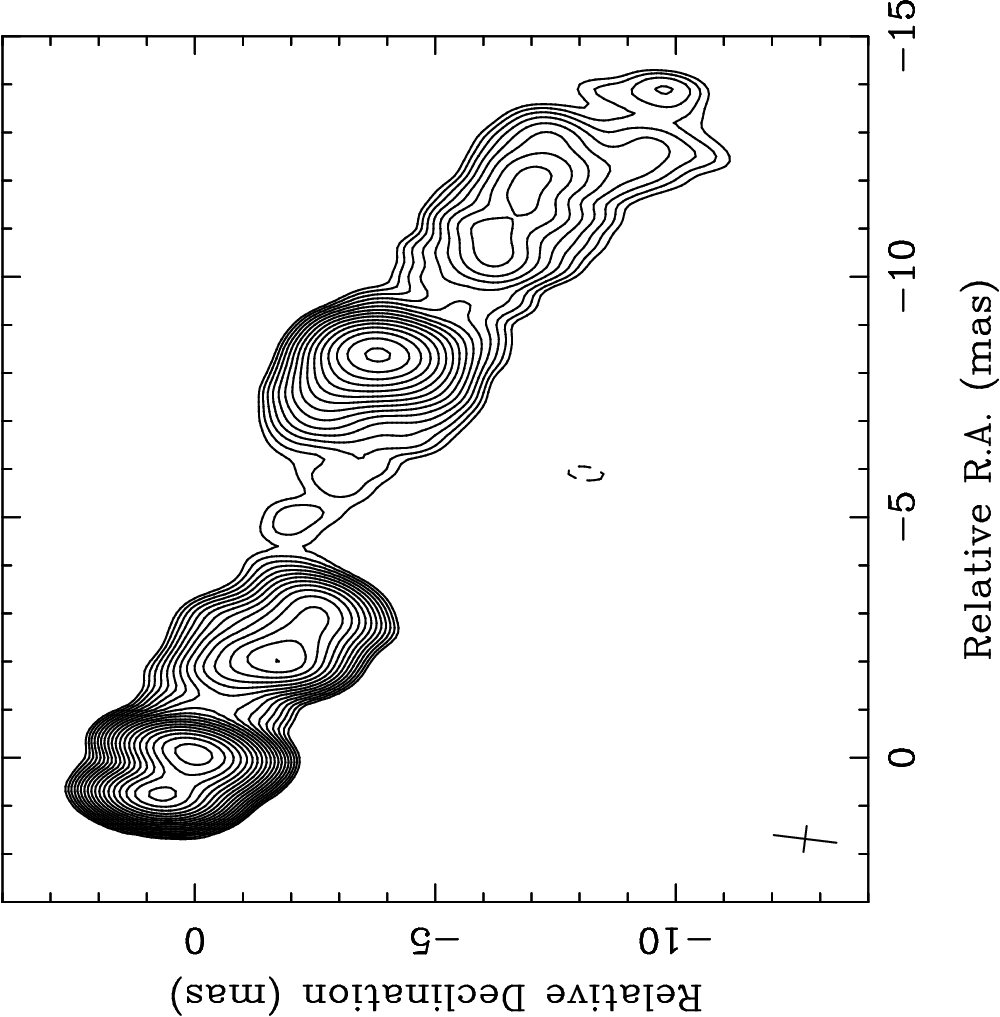}}
\resizebox{0.70\hsize}{!}{\includegraphics[angle=-90,clip=true]{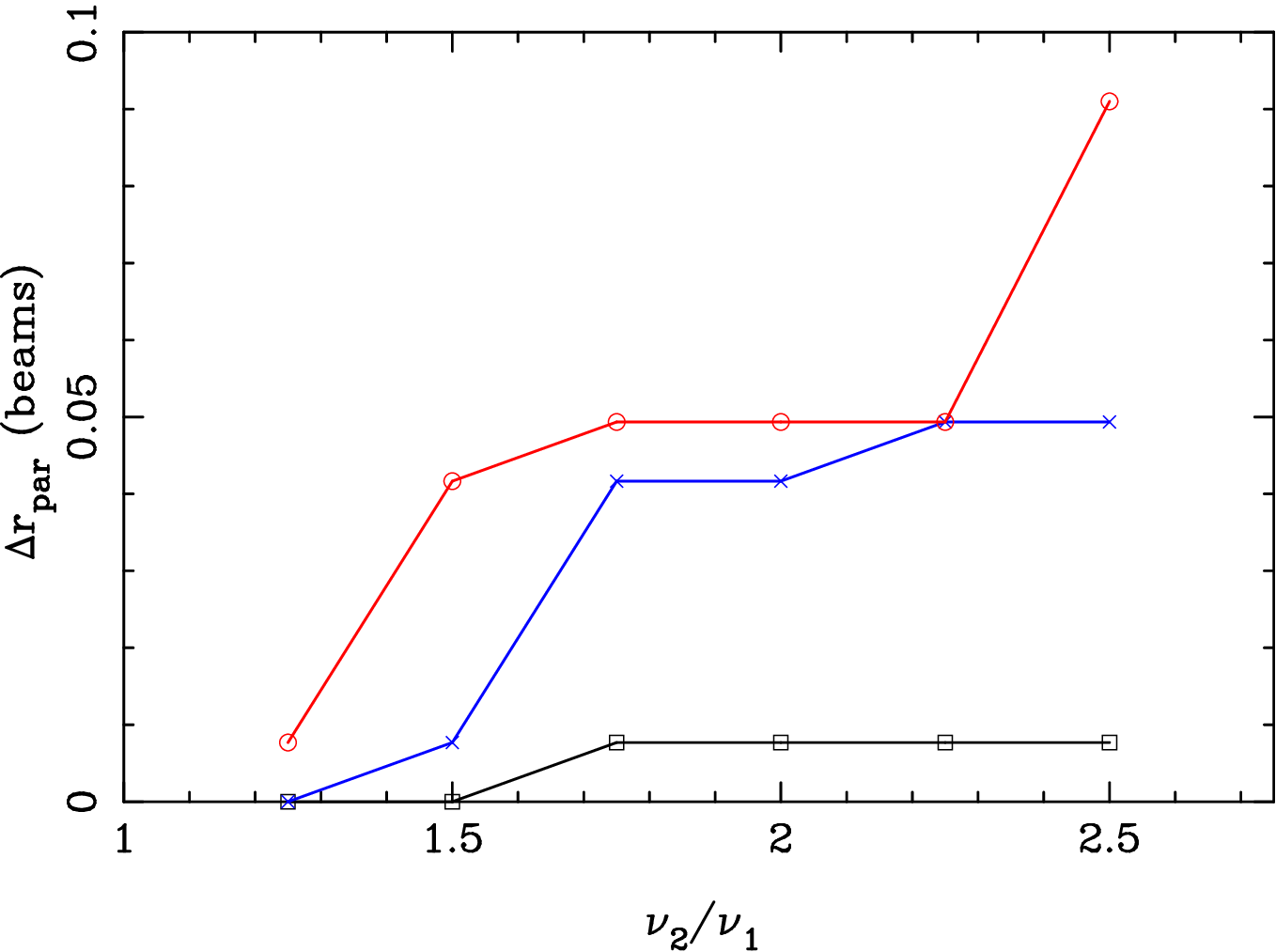}}
\resizebox{0.70\hsize}{!}{\includegraphics[angle=-90,clip=true]{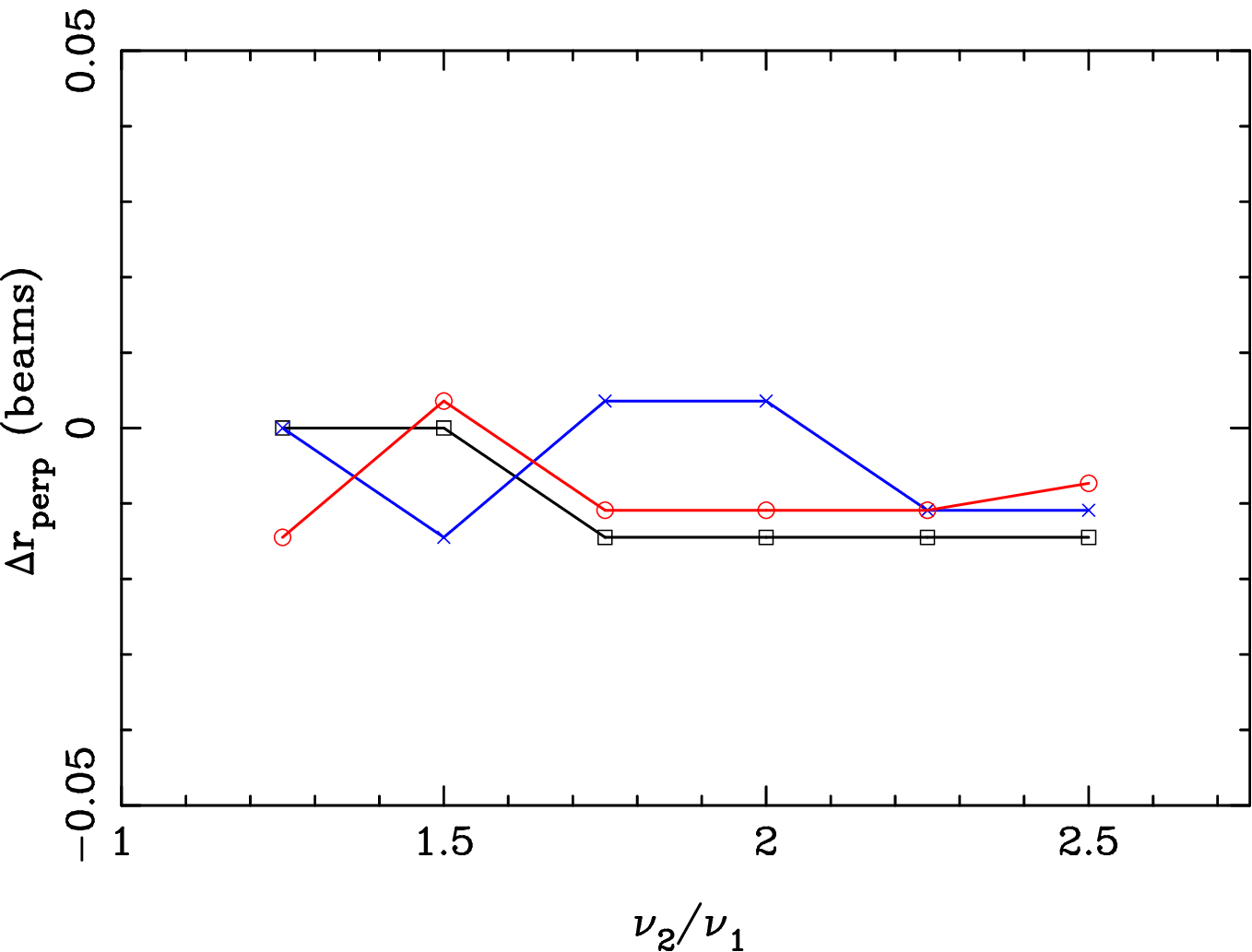}}
\end{center}
 \caption{ Naturally weighted total intensity CLEAN image of 3C273 at 15.4~GHz (top). 
           The alignment error (in beam widths) along the jet as a function of 
           frequency ratio (middle). The alignment error (in beam widths) perpendicular 
           to the jet (bottom). The error is defined as $\vec{r_{\nu_1}}-\vec{r_{\nu_2}}$ 
           where $\vec{r_{\nu_1}}$ is the position in the low-frequency image and 
           $\vec{r_{\nu_2}}$ is the corresponding position in the high-frequency image. 
           The different spectral index gradients are shown in different colors: 
           $-$0.1~mas$^{-1}$ (black squares), $-$0.2~mas$^{-1}$ (blue crosses), and $-$0.3~mas$^{-1}$ (red circles). }
 \label{f:3C273}
\end{figure}

\subsection{Systematic errors}
So far we have discussed random errors, but the image
registration method is also prone to systematic errors that may bias
our core shift measurements. Namely, the assumption of a constant
spectral index across the matched features does not necessarily hold
in real jets. Indeed, it is known that spectral index gradients
occur along the jet. The spectral index images between 8.1 and
15.4~GHz typically show gradients in the optically thin part of the
jet from $-0.3$~mas$^{-1}$ to $+0.1$~mas$^{-1}$ with an average of
$\sim-0.1$~mas$^{-1}$ (Hovatta et al., in prep.).

We tested the systematic effect that such gradients may have on
image registration results by performing simulations. For the
simulations we selected five sources with different jet morphologies:
NGC\,315 (straight, long and very smooth jet), 0716+714 (straight,
short and smooth jet; Fig.~\ref{f:0716+714}, top), 3C\,273 (long,
wiggling jet with prominent knots; Fig.~\ref{f:3C273}, top), BL\,Lac
(wide, curved jet having one prominent knot downstream of the core),
and 3C\,454.3 (complex, curved jet with knots). For each source we
created three sets of simulated images at six different frequencies
exhibiting three different spectral index gradients along the jet:
$-0.1$, $-0.2$, and $-0.3$~mas$^{-1}$. The simulated images were based
on a real 15.4~GHz image of a given source, to which a constant spectral
index gradient along the jet was applied and new images at frequencies
of 1.25, 1.50, 1.75, 2.00, 2.25, and 2.50 times the original frequency
were calculated. Finally, random noise at the same level as in the
original image was added to simulated images to ensure that
background noise patterns between the images do not correlate. The
original image and the simulated one were then registered using
NCC. Note that this simulation setup provides a worst-case scenario
in the sense that the gradient is assumed to be constant for the whole
jet length, whereas in the real jets this is typically not the case.

The simulation results show that spectral index gradients along
the jet can indeed affect the registration results and that the
systematic error introduced this way depends on the jet morphology,
the magnitude of the gradient, and the frequency ratio of the
registered pair of images. In 3C\,273 and BL\,Lac, which have
significant structural detail in the optically thin part of the jet,
the errors in the registered shift along the jet are less than 5\%
and less than 3\% of the beam size, respectively, for all values
of the gradient and for frequency ratios lower than 2. For 3C\,273, 
a gradient of $-0.1$~mas$^{-1}$ results in an error that is less 
than 1\% of the beam size (Fig.~\ref{f:3C273}, middle). The jet in 
3C\,454.3 also has significant structural detail, and the systematic 
errors for a gradient of $-0.1$~mas$^{-1}$ stay below 4\% of the 
beam size. However, for steeper gradients, the errors increase 
significantly, being less than $\sim$10\% of the beam size for frequency 
ratios below 2. The featureless jets of NGC\,315 and 0716+714 are the 
most prone to systematic errors caused by spectral index gradients along 
the jet: even the flattest gradient results in errors in the range of
10--18\% of the beam size. In 0716+714 the gradients of $-0.2$ and
$-0.3$~mas$^{-1}$ result in errors of 24\% and 27\% of the beam size,
respectively, at a frequency ratio of 2 (Fig.~\ref{f:0716+714}, middle). 
In NGC\,315 the steeper gradients cause a significant jump in the 
image shift that exceeds the beam size.

The above simulation demonstrates that the level of detail in
the optically thin part of the jet is crucial for the reliability of
the cross-correlation based image registration. If the jet has knots
or bends, cross-correlation is rather robust against possible
spectral index gradients. On the other hand, the method clearly does 
not work with smooth, straight jets that exhibit spectral
index gradients. Also, the direction of the erroneous image shift
due to spectral index gradient along these smooth, straight jets is
such that it can mimic a true core shift. Therefore, we have
repeated the core shift analysis in a ``clean sample'' from which
such jets are removed. Statistics on the clean sample that comprises
94 sources, however, did not show lower median values for 15.4-8.1, 
15.4-8.4, and 15.4-12,1~GHz core shift distributions. This indicates 
that the effect of a possible overestimation of image shifts (and 
consequently core shifts) owing to a spectral index gradient along 
the jet is weak.
%The clean sample comprises 94 sources. It turned out, that the excluded
%sources are characterized by statistically smaller shifts, so that the
%15.4-8.1, 15.4-8.4, and 15.4-12.1~GHz core shift distributions on the
%clean sample have the medians $\sim$10\% higher than for the whole sample.
%Therefore, even if some short-straight-smooth jets have overestimated image
%shifts (and consequently core shifts) due to spectral index gradient, this
%effect is weaker comparing to a tendency for longer jets to have on average
%larger core shifts. Indeed, the larger viewing angle (longer jet) the larger
%core shift is expected and vice versa.

We also estimated positional errors (from the image plane) of 
the core components using the relation suggested by \cite{F99}: 
$(W \cdot RMS)/(2P)$, where $W$ and $P$ are the convolved size and 
peak intensity of the component, $RMS$ is the post-fit r.m.s. error 
associated with the pixels in a nine-beam area region under the 
component in the residual image. Because the cores are bright and 
compact, their signal-to-noise ratio $S/N=P/RMS$ values are high, 
with a median value of a few hundred, making the corresponding 
positional errors small, with a median value of about $~1$~$\mu$as. 
This implies that the core shift error is typically dominated by 
the uncertainty of the image alignment for a sample of core-dominated 
AGN jets.
%It should also be noted that the core components are not isolated, suggesting that real position errors might 
%be larger. Nevertheless, even if the position errors are underestimated by a factor of 10, it does not change the 
%conclusion of dominating contribution to the core shift uncertainty coming from a 2D cross-correlation technique.

The typical accuracy level of about 50~$\mu$as achieved in our core
shift measurements is comparable to that from the self-referencing
approach \citep{Sokolovsky_11cs}, where the errors are dominated by
position uncertainties of a referencing jet component, which is
usually larger in size and weaker in flux density with respect to the
core. Phase-referencing observations may provide slightly better
accuracy, down to 30~$\mu$as, for the 15.4-8.4~GHz core shift, as
reported by \cite{Hada_M87}, but cannot be used for a large number of
sources as discussed in Sect.~\ref{intro}. Another complication of
the phase-referencing method comes from the fact that the calibrator
has its own core shift.

\subsection{Stationary jet feature problem}
Jets can exhibit standing features, like re-collimation shocks, in or 
close to their core region and these features could in principle 
introduce a level of artificial core shift owing to the degradation 
of the angular resolution with wavelength. To study how strong an 
effect these standing features could have, we used the AIPS task 
UVMOD to create simulated VLBA data sets at 8.1, 8.4, 15.3, 23.8, 
and 43.2~GHz. We used a real multi-wavelength VLBA observation to
provide the $(u,v)$ plane sampling and noise properties at every 
frequency, and as an input model for UVMOD we used three Gaussian 
components representing the core, a standing feature, and a jet 
feature. A set of simulated $(u,v)$ data was generated with different 
flux ratios between the core and the standing feature (ranging from 
10\% to 50\%) and with different distances between the two (0.15~mas 
and 0.3~mas). As a result, we had a simulated multi-wavelength VLBA
data set in which all wavelengths had exactly the same sky brightness 
distribution, but different $(u,v)$ sampling and noise. Any apparent 
core-shift in this data set would be purely caused by the $(u,v)$ 
coverage differences at different frequencies. Analyzing the simulated 
data sets, we found that a close (within 0.3~mas from the 43~GHz core) standing jet feature can contribute to 
the expected core shift effect between 43~GHz and 8~GHz owing to 
frequency-dependent blending on the level of $\sim$10\% of the expected 
core shift if the flux density ratio of the stationary feature and the 
core $S_\mathrm{st}/S_\mathrm{c}\sim10$\%, and reaching up to $\sim$30\%
 of the expected core shift if $S_\mathrm{st}/S_\mathrm{c}\sim50$\%. 
Between the frequencies used in this study, i.e., 15 and 8~GHz, the 
core shift is affected by the frequency-dependent blending by 
$\lesssim 20$~$\mu$as, which is less than the estimated accuracy of 
our measurements. We note that these simulations describe a very simple 
situation and a more detailed study of the effect of stationary features 
to the measured core shift using real data is warranted. However, this 
is beyond the scope of the current work.

\section{Jet physics with core shifts}

\begin{figure}
\begin{center}
%\resizebox{0.49\hsize}{!}{\includegraphics[angle=-90,clip=true]{FIGS/fig07a.eps}}
%\resizebox{0.50\hsize}{!}{\includegraphics[angle=-90,clip=true]{FIGS/fig07b.eps}}
\resizebox{0.49\hsize}{!}{\includegraphics[angle=-90,clip=true]{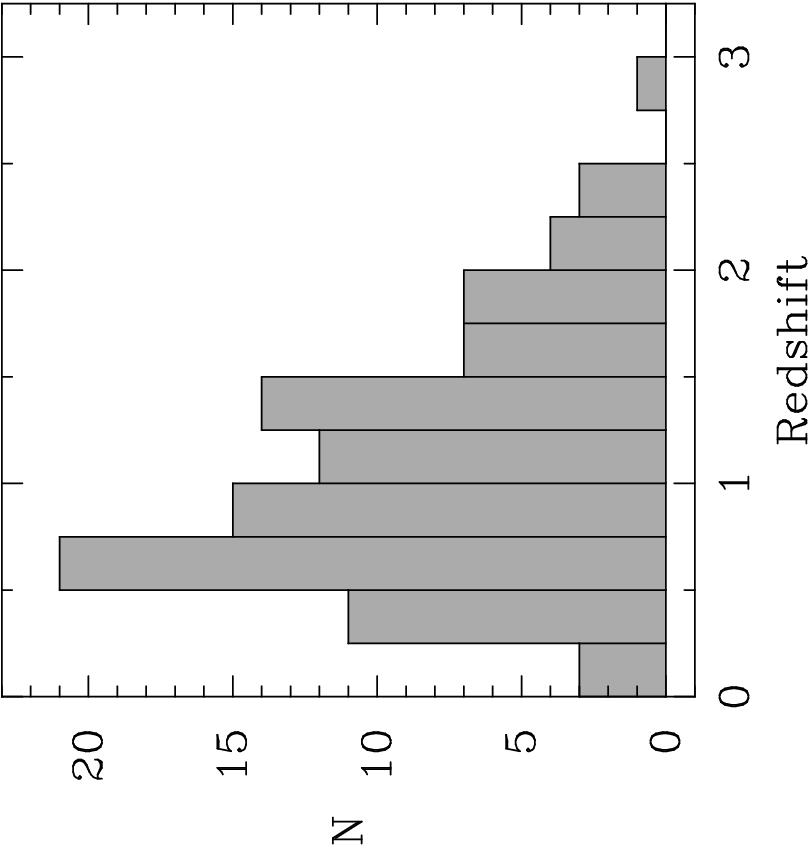}}
\resizebox{0.50\hsize}{!}{\includegraphics[angle=-90,clip=true]{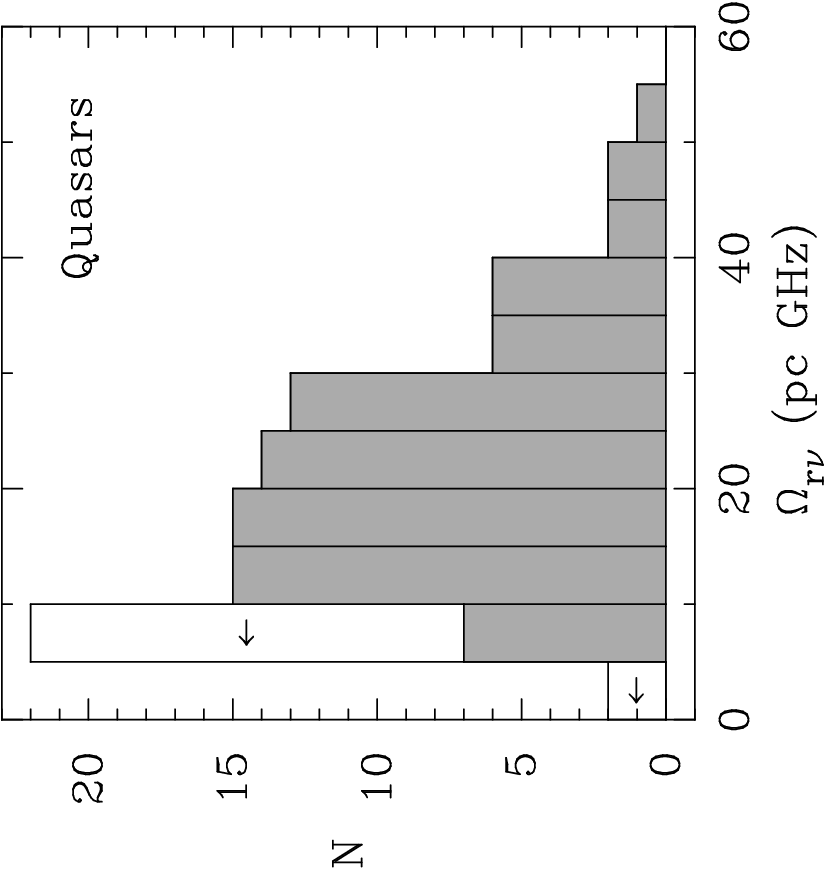}}
\resizebox{0.49\hsize}{!}{\includegraphics[angle=-90,clip=true]{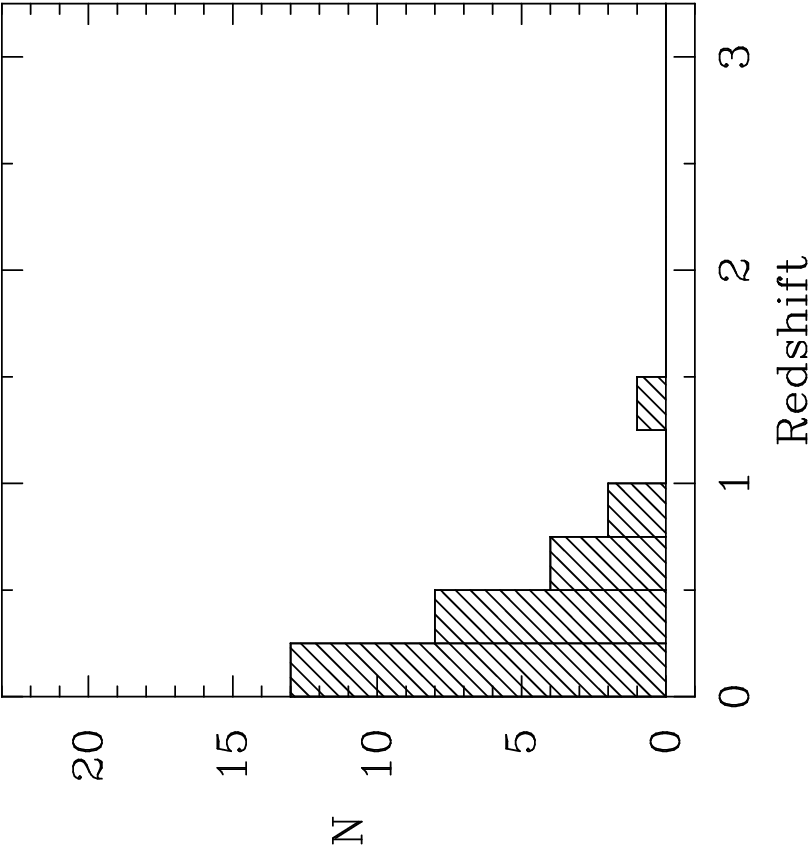}}
\resizebox{0.50\hsize}{!}{\includegraphics[angle=-90,clip=true]{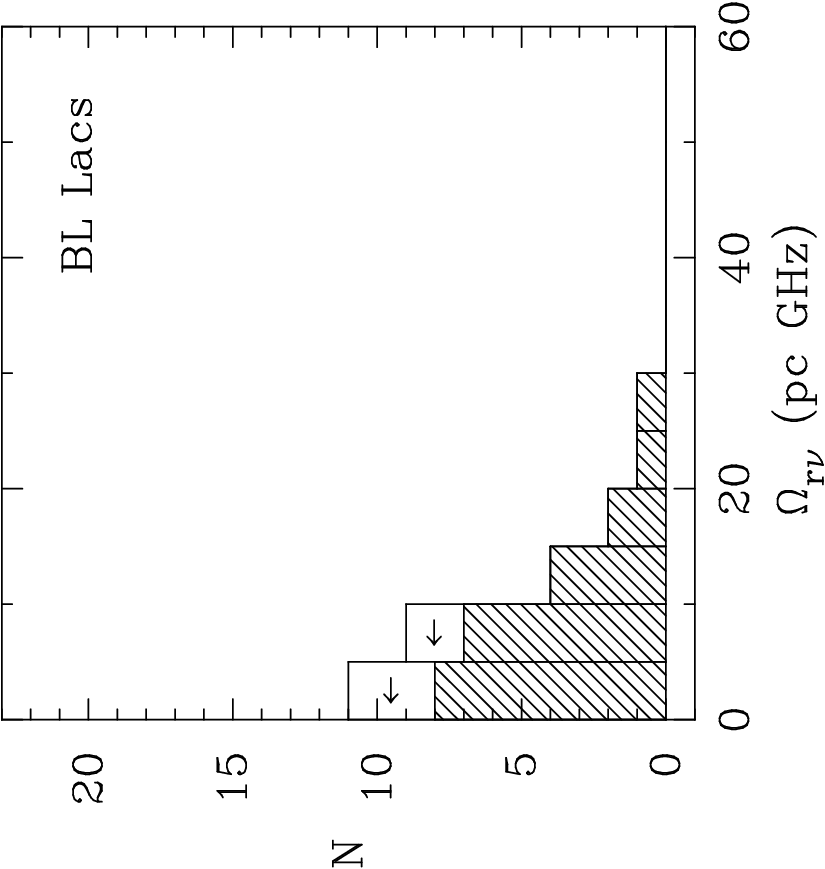}}
\resizebox{0.49\hsize}{!}{\includegraphics[angle=-90,clip=true]{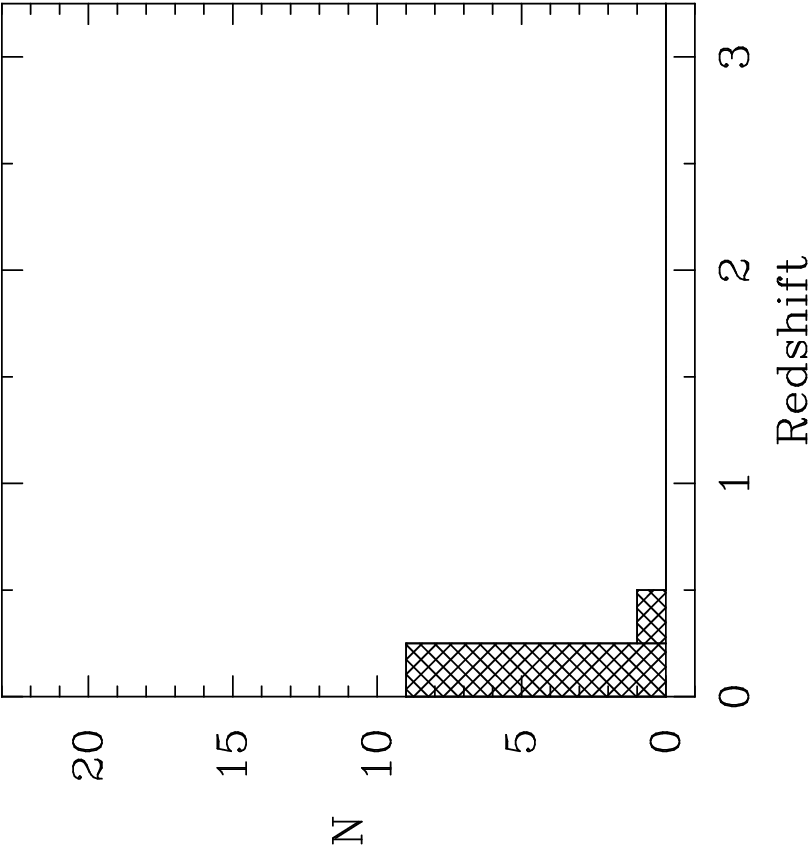}}
\resizebox{0.50\hsize}{!}{\includegraphics[angle=-90,clip=true]{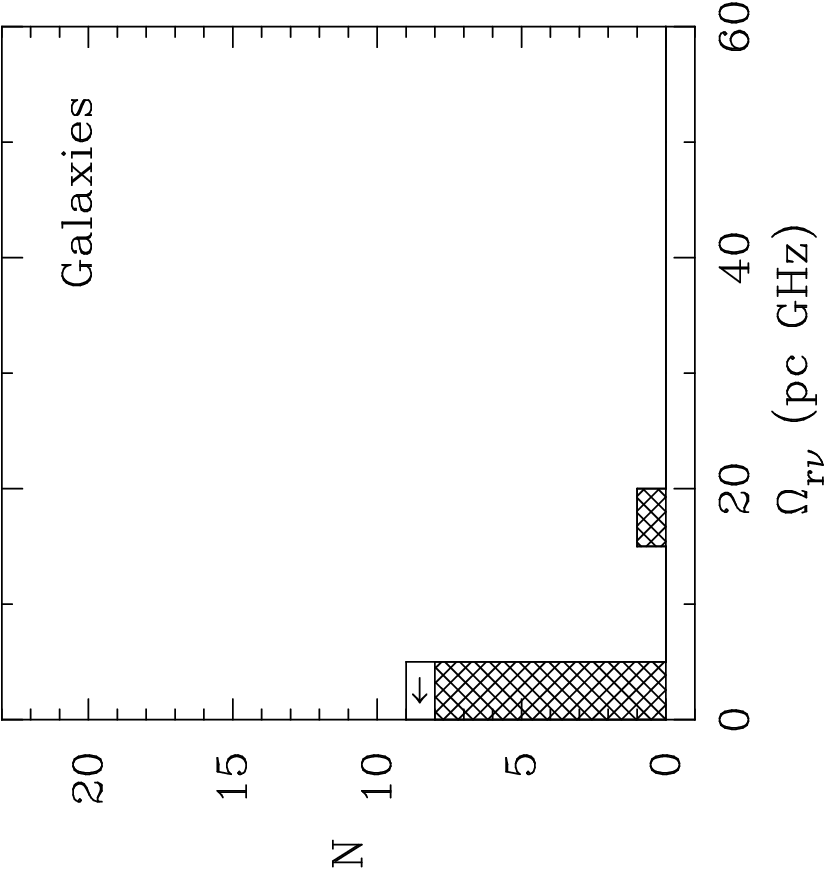}}
\end{center}
 \caption{ Distribution of redshift (left) and core shift measure
   $\Omega_\mathrm{r\nu}$ (right) for 136 sources with the 
   core shift derived between 15 and 8~GHz. Gray filled bins represent 98
   quasars (top), dashed bins represent 28 BL~Lacs (middle), and cross-hatched bins represent
   10 galaxies (bottom). Empty bins show upper limits. }
 \label{f:z_omega_hist}
\end{figure}

To increase the robustness of our subsequent calculations, from now on 
we use a vector average for each pair of the 15.4-8.1 and 15.4-8.4~GHz 
core shifts. Moreover, if the core shift magnitude is smaller than 1 
sigma level, we set $\Delta r_\mathrm{core\,15-8}=\sigma_{15-8}=51$~$\mu$as 
to be used as the upper limit, since eliminating the small core shifts 
would introduce a bias. We excluded only the core shift vectors with
$|\theta_\mathrm{cs}-\theta_\mathrm{jet}|>90\degr$, which are likely
to be dominated by errors. In total, we have core shifts between 15 and 
8~GHz for 136 sources, 108 of which have both redshift 
(Fig.~\ref{f:z_omega_hist}, left) and apparent jet speed measurements.
If a source had a second epoch, we selected the one at which the 
dynamic range of the 15 GHz image was higher.

As shown by \cite{L98}, core shift measurements can be used for
deriving a variety of physical conditions in the compact jets. In
particular, assuming equipartition between the particle and magnetic
field energy density ($k_r=1$) and a jet spectral index
$\alpha_\mathrm{jet}=-0.5$ ($S\propto\nu^{+\alpha}$), the magnetic
field in Gauss at 1~pc of actual distance from the jet vertex can be calculated
through the following proportionality \citep{Hirotani05,Sullivan_09cs}
\begin{equation}
B_1\simeq0.025\left(\frac{\Omega^3_{r\nu}\,(1+z)^2}{\varphi\,\delta^2\sin^2\theta}\right)^{1/4}\,,
\label{e:b1}
\end{equation}
where $\varphi$ is the half jet opening angle, $\theta$ is the viewing
angle, $\delta$ is the Doppler factor, and $\Omega_{r\nu}$ is the core
shift measure defined in \cite{L98} as
\begin{equation}
\Omega_{r\nu}=4.85\cdot10^{-9}\,\frac{\Delta r_\mathrm{core,\,\nu_1\nu_2}\,D_L}{(1+z)^2}\cdot\frac{\nu_1\nu_2}{\nu_2-\nu_1}\ \mathrm{pc \cdot GHz}\,,
\end{equation}
where $\Delta r_\mathrm{core,\,\nu_1\nu_2}$ is the core shift in
milliarcseconds, and $D_L$ is the luminosity distance in parsecs. 
The calculated values of $\Omega_{r\nu}$ in pc$\,\cdot\,$GHz form a
distribution ranging from 0.8 to 54.1 and peaking near the median of
13.6 (Fig.~\ref{f:z_omega_hist}, right). The distributions of
$\Omega_\mathrm{r\nu}$ for quasars and BL~Lacs are significantly
different ($p<10^{-4}$) as indicated by Gehan's generalized 
Wilcoxon test from the ASURV survival analysis package \citep{ASURV}, 
with medians of 18.6 and 7.1 pc$\,\cdot\,$GHz, respectively.

\begin{figure}
\begin{center}
\resizebox{0.99\hsize}{!}{\includegraphics[angle=-90,clip=true]{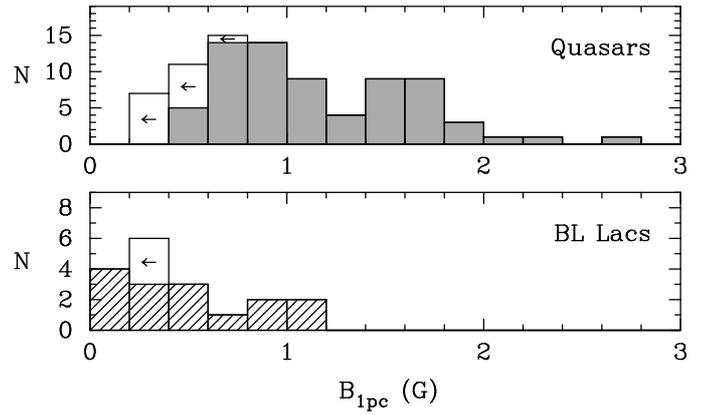}}
\end{center}
 \caption{ Distribution of magnetic field at a distance of 1~pc from
   the black hole for 84 quasars (top), 18 BL~Lacs (bottom). Empty bins 
   represent upper limits. }
 \label{f:b1}
\end{figure}

The combination of $\varphi$, $\delta$, and $\theta$ in Eq.~(\ref{e:b1}) 
is typically known for only a small fraction of sources, limiting the 
applicability of the formula. Therefore, the number of sources in our 
subsample with known apparent jet speed $\beta_\mathrm{app}$ 
\citep{MOJAVE} is larger by a factor of $>$2 than the number of 
sources with known variability Doppler factor \citep{Hovatta09}, 
intrinsic opening angle and viewing angle \citep[e.g.][]{Pushkarev09,MF4}.  
The denominator in (\ref{e:b1}) can be expressed through 
$\beta_\mathrm{app}$ only by substituting
$\delta=\Gamma^{-1}(1-\beta\cos\theta)^{-1}$ and
%lower bound 
$\Gamma_\mathrm{min}=(1+\beta^2_\mathrm{app})^{1/2}$, and also taking
into account that $2\varphi\simeq0.26\,\Gamma^{-1}$
\citep{Pushkarev09} and
$\beta_\mathrm{app}=\beta\sin\theta\,(1-\beta\cos\theta)^{-1}$. With
these substitutions we are assuming the the jet is viewed at the critical
angle $\theta\simeq\Gamma^{-1}$ that maximizes
$\beta_\mathrm{app}$. We therefore obtain
\begin{equation}
B_1\simeq0.042\,\Omega^{3/4}_{r\nu}\,(1+z)^{1/2}(1+\beta_\mathrm{app}^2)^{1/8}\,,
\label{e:b1_mod}
\end{equation}
where for $\beta_\mathrm{app}$ we use the fastest non-accelerating,
radial apparent speed measured in the source \citep{MOJAVE}.

First, we tested the consistency of $B_1$ values calculated from
Eqs.~(\ref{e:b1}) and (\ref{e:b1_mod}) for sources with previously
measured $\delta$ and $\beta_\mathrm{app}$ values. For 40 sources out
of 43 in common, comprising 
% I tossed out 0814+425 as it has no published z 
35 quasars and 8 BL Lacs, the results agree within the errors, with a
median value of their ratio of 0.99. For three sources (0420$-$014,
0804+499, and 1413+135), Eq.~(\ref{e:b1}) gives several times higher
values, most probably due to underestimated apparent speeds, to which
(\ref{e:b1}) is more sensitive than (\ref{e:b1_mod}), because the
viewing angle
$\theta=\arctan[2\beta_\mathrm{app}(\beta_\mathrm{app}^2+\delta^2-1)^{-1}]$
and the opening angle
$\varphi=\varphi_\mathrm{obs}\sin\theta$. Indeed, these sources have
low apparent speeds but high Doppler factors, leading to low viewing
angle estimates of $1\fdg9$, $0\fdg2$, $1\fdg4$ \citep{MF4},
respectively, and in turn to overestimated magnetic field
strengths. The uncertainties in $B_1$ from Eq.~\ref{e:b1_mod} were
calculated taking into account the errors in the core shifts, apparent
speeds, and also from the assumption $\theta\simeq\Gamma^{-1}$, which 
is known to introduce an additional source of errors, as discussed by
\cite{Lister_thesis}. Since the aforementioned assumption is less
correct for sources with low Lorentz factors, we excluded galaxies from
the subsequent analysis.

\begin{table*}
\caption{Jet parameters derived for the sources with significant core shifts between 8 and 15~GHz.}
\label{t:b_field}
\centering
\begin{tabular}{c c c c c r r r r r} \hline\hline

  Source   &      Epoch &Opt. class& $z$ & $\beta_\mathrm{app}$ & $\Delta r_\mathrm{core,\,15-8\,GHz}$ & $\Omega_{r\nu}$ & $B_1$ & $B_\mathrm{core}$ & $r_\mathrm{core,\,15.4\,GHz}$ \\
           &            &          &     &                      &                         (mas) &        (pc GHz) &   (G) &               (G) &                  (pc) \\
\hline
0003$-$066 & 2006-07-07 & B & 0.347 &  2.89 & $<$0.051 & $<$4.55 & $<$0.20 & $>$0.22 & $<$ 0.91\\
0106$+$013 & 2006-07-07 & Q & 2.099 & 26.50 & $<$0.051 & $<$7.86 & $<$0.79 & $>$0.06 & $<$13.56\\
0119$+$115 & 2006-06-15 & Q & 0.570 & 17.10 &    0.324 &   37.75 &    1.63 &    0.04 &    42.07\\
0133$+$476 & 2006-08-09 & Q & 0.859 & 12.98 &    0.099 &   13.69 &    0.77 &    0.07 &    11.60\\
0149$+$218 & 2006-02-12 & Q & 1.320 & 18.55 &    0.196 &   29.64 &    1.69 &    0.05 &    35.83\\
0202$+$149 & 2006-09-06 & Q & 0.405 &  6.41 &    0.113 &   10.87 &    0.48 &    0.10 &     4.59\\
0212$+$735 & 2006-07-07 & Q & 2.367 &  7.64 &    0.143 &   21.24 &    1.27 &    0.12 &    10.65\\
0215$+$015 & 2006-04-28 & Q & 1.715 & 34.16 &    0.111 &   17.08 &    1.41 &    0.04 &    37.98\\
0224$+$671 & 2006-10-06 & Q & 0.523 & 11.63 &    0.139 &   15.47 &    0.75 &    0.06 &    11.75\\
0234$+$285 & 2006-09-06 & Q & 1.207 & 12.26 &    0.239 &   35.82 &    1.71 &    0.06 &    28.67\\
\hline
\end{tabular}
\tablefoot{
Apparent speed values $\beta_\mathrm{app}$ are taken from \cite{MOJAVE}.
Table \ref{t:b_field} is published in its entirety in the electronic version of 
{\it Astronomy \& Astrophysics}. A portion is shown here for guidance regarding
its form and content.}
\end{table*}

\begin{figure}
\begin{center}
\resizebox{0.99\hsize}{!}{\includegraphics[angle=-90,clip=true]{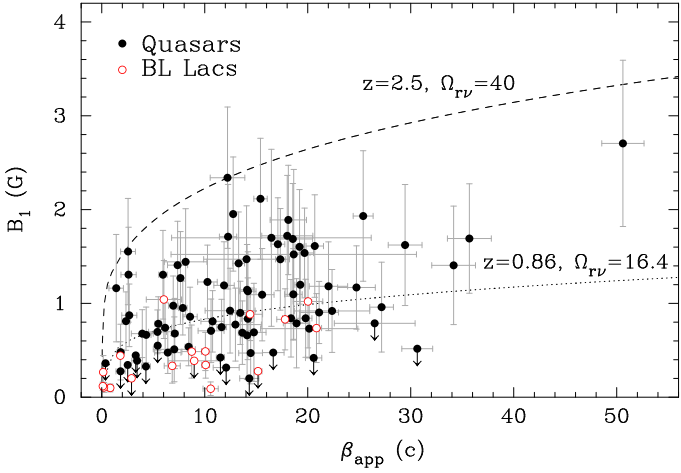}}
\end{center}
 \caption{ Magnetic field at a distance of 1~pc from the central black
   hole versus fastest non-accelerating, radial apparent speed. The
   measurements are enveloped under the dashed aspect line. The middle
   dotted line shows the dependence based on the medians for redshift
   and core shift measure $\Omega_{r\nu}$.  }
 \label{f:b1_vs_beta}
\end{figure}

The distributions of $B_1$ values derived from Eq.~(\ref{e:b1_mod})
for 84 quasars and 18 BL~Lacs shown in Fig.~\ref{f:b1} have medians of
%$0.91^{+0.25}_{-0.09}$ and $0.43^{+0.29}_{-0.12}$~G, 
$0.9^{+0.2}_{-0.1}$ and $0.4^{+0.3}_{-0.1}$~G, respectively, where
the errors derived from a bootstrapping method indicate 95\% confidence
intervals. Gehan's generalized Wilcoxon test from the ASURV 
survival analysis package indicates that the distributions of $B_1$ 
for quasars and BL~Lacs are different at 99.6\% confidence level.
%A K-S test indicates that the distributions of $B_1$ for quasars and 
%BL~Lacs are different at 99.5\% confidence level. 
The difference is driven mainly by statistically higher redshifts for the 
quasars, and, though to a lesser degree, also by higher $\beta_\mathrm{app}$
(Fig.~\ref{f:b1_vs_beta}). The $B_1$ values would be comparable for
these classes of objects, if the ratio of the median core shifts for
BL~Lacs to that of quasars was about 2.7, but that is not the
case. Systematically stronger magnetic fields in quasars can be a
result of more massive black holes hosted by them and/or higher
accretion rate in these objects, bacause larger black holes can accrete
more matter, effectively powering the jets and accelerating particles
to higher speeds. This scenario is also supported by an indication for
quasars to have on average narrower intrinsic opening angles than
those of BL~Lacs, as reported by \cite{Pushkarev09}. At the same time,
\cite{Woo02} argued that BL~Lacs host black holes of comparable mass to 
quasars, though the estimates of black hole masses for quasars and
BL~Lacertae objects have been generated by different methods, and this
may introduce a bias while comparing the mass assessments.

\begin{figure}
\begin{center}
\resizebox{0.99\hsize}{!}{\includegraphics[angle=-90,clip=true]{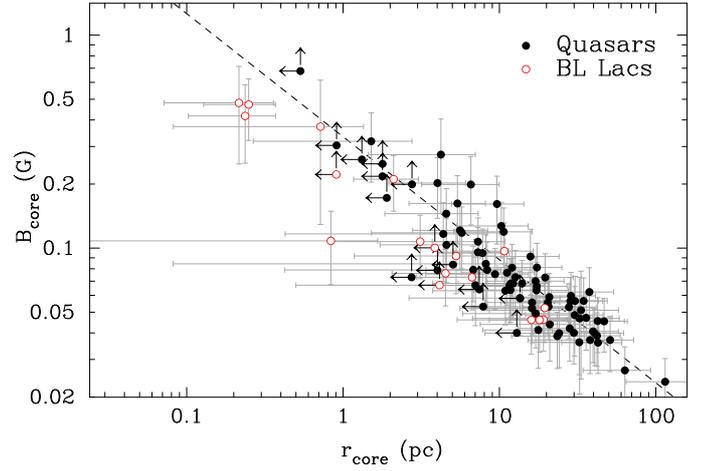}}
\end{center}
 \caption{ Magnetic field in the 15~GHz core versus its deprojected
   separation from the true jet base. The dashed line is the linear
   least-squares fit for quasar measurements.
   }
 \label{f:b_core_vs_r_core}
\end{figure}

The absolute distance in parsecs of the apparent VLBI core measured
from the jet vertex is given by \citep{L98}
\begin{equation}
r_\mathrm{core}(\nu)=\frac{\Omega_{r\nu}}{\nu\sin\theta}\approx\frac{\Omega_{r\nu}(1+\beta^2_\mathrm{app})^{1/2}}{\nu}\,,
\end{equation} 
where $\nu$ is the observed frequency in GHz. Then the corresponding
magnetic field strength is
$B_\mathrm{core}=B_1r_\mathrm{core}^{-1}$. In
Fig.~\ref{f:b_core_vs_r_core}, we plot the derived $r_\mathrm{core}$
and $B_\mathrm{core}$ for the 15~GHz cores. Note that the core
magnetic fields show a reverse tendency compared to the $B_1$ values,
with a median of 0.10~G for BL~Lacs and 0.07~G for quasars, because
their apparent cores are at different separations from the true jet
base, with medians of 4.0, and 13.2~pc, respectively.

On the projected plane (i.e., a VLBI map), the median separation of
the 15~GHz core from the central black hole is about 1~pc,
corresponding to about 0.14~mas for a source at $z\simeq1$. We
summarize the derived physical quantities, such as $\Omega_{r\nu}$,
$B_1$, $B_\mathrm{core}$, and $r_\mathrm{core}$ in
Table~\ref{t:b_field}.

The mass of a central black hole can be related to $B_1$, such that
$M_{bh}\approx2.7\times10^9B_1M_{\sun}$ \citep{L98}.  Its typical value
is $\sim$$10^9M_{\sun}$ for quasars and BL~Lacs. 
$B_1$ can also be used to estimate the magnetic field $B_\mathrm{BH}$ 
near the central black hole, assuming a $B\propto R^{-1}$ dependence, 
where $R$ is the half-width of the jet. Thus, we have 
$B_\mathrm{BH}=B_1R_1/R_\mathrm{BH}$, where $R_1$ and $R_\mathrm{BH}$ 
are the the half-widths of the jet at 1~pc from the jet vertex and near 
the black hole, respectively. To derive $R_1$, we assume that the jet 
is conical at distances larger than 1~pc. Then, we have 
$R_1 = R_\mathrm{core} - (r_\mathrm{core} - 1)\tan\varphi$,
where $R_\mathrm{core}$ is the half-width of the 15~GHz core, 
$r_\mathrm{core}$ is the 15~GHz core separation from the jet vertex 
(both measured in pc), and $\varphi$ is the intrinsic half jet opening 
angle. Because the black hole masses for BL~Lacs are poorly known due to 
their weak emission lines, we restrict our analysis to quasars only. 
We derived the following median values: $\varphi = 0\fdg6$ 
\citep{Pushkarev09}, $r_\mathrm{core} = 13.2$~pc, 
$R_\mathrm{core} = 0.35$~pc. Then $R_1 \approx 0.2$~pc and 
$B_\mathrm{BH} \approx 2\times10^{3}$~G, assuming the jet width near 
the $\sim$$10^9M_{\sun}$ black hole to be on the order of its gravitational 
radius. The derived assessment of the magnetic field near the black 
hole is consistent with the theoretical value calculated from a model 
of a thin, magnetically driven accretion disk \citep{Field93}.
%The magnetic field in the jet-launching region situated at a distance 
%of $\sim$$100R_g$ \citep{Lobanov_07} would then be $\sim$$10^2$~G.

\section{Summary}
\label{s:summary}

We have implemented a method for measuring the frequency-dependent shift
in absolute position of the parsec-scale core and applied it to
multi-frequency (8.1, 8.4, 12.1, and 15.4~GHz) VLBA observations of
191 sources performed during 2006 within the MOJAVE program. The
method is based on results from (i) image registration achieved
by a two-dimensional cross-correlation technique and (ii) structure model
fitting. It has proved to be very effective and provided the core
shifts in 163 sources (85\%), with a median of 128~$\mu$as between 15
and 8~GHz, and 88~$\mu$as between 15 and 12~GHz. Despite the moderate
separation of the observing frequencies, the derived core shifts are
significant ($>$$2\sigma$) in about 55\% of cases, given an estimated typical
uncertainty of 50~$\mu$as between 15 and 8~GHz, and 35~$\mu$as between
15 and 12~GHz. The errors are dominated by uncertainties from the two-dimentional
cross-correlation procedure, because the relative positional uncertainties
of the compact bright cores are at a level of a few
microarcseconds. The significant core shift vectors are found to be
preferentially aligned with the median jet direction, departing from
it by less than $30\degr$ in more than 90\% of cases.  
%We have also found an indication for the angular core shift to decrease with increasing redshift.

We used the measured core shifts for constraining magnetic field
strengths and core sizes for 89 sources. The magnetic field at a
distance of 1~pc from the jet injection point is found to be $\sim$0.9~G
for quasars and $\sim$0.4~G for BL~Lacs. Extrapolating all the way
back by assuming a $B\propto R^{-1}$ dependence, the magnetic field in
%the jet apex of blazars located at $\sim$$100R_g$ is about $10^2$~G, and in 
the close vicinity of the black hole
%, at $\sim$$1R_g$ 
is about $2\times10^3$~G. The core sizes, i.e., the distances 
from the true jet base to its apparent origin at 15.4~GHz are statistically 
larger in quasars than in BL~Lacs, with a median of 13.2 and 4.0~pc,
respectively.  At these distances, the magnetic field has a median of
0.07~G for quasars and 0.1~G for BL~Lacertae objects.

Future multi-epoch and multi-frequency VLBI observations (including
phase-referencing) of a pre-selected sample of sources with prominent 
core shifts are needed to address the question of the core shift variability, 
which can be used not only for astrophysical studies but also for astrometric
applications. Ideally, these observations should be performed during and
after strong nuclear flares that can be detected in advance in higher
energy domains, e.g., optical or gamma-ray, and cover a wide range of
observing frequencies, extending down to 5 or 2~GHz, where synchrotron
self-absorption is essential.

\begin{acknowledgements}

We would like to thank K.~I. Kellermann, E. Ros, D.~C. Homan and the 
rest of the MOJAVE team for the productive discussions. We thank the 
anonymous referee for useful comments, which helped to improve the 
manuscript. This research has made use of data from the MOJAVE database 
that is maintained by the MOJAVE team \citep{MOJAVE}. The MOJAVE project 
is supported under National Science Foundation grant AST-0807860 and NASA 
{\it Fermi} grant NNX08AV67G. T.H. was supported in part by the Jenny and 
Antti Wihuri foundation. YYK is partly supported by the Russian Foundation 
for Basic Research (project 11-02-00368), Dynasty Foundation, and the 
basic research program ``Active processes in galactic and extragalactic 
objects'' of the Physical Sciences Division of the Russian Academy of 
Sciences. Part of this work was supported by the COST Action MP0905 
``Black Holes in a Violent Universe''. The VLBA is a facility of the 
National Science Foundation operated by the National Radio Astronomy 
Observatory under cooperative agreement with Associated Universities, Inc.

\end{acknowledgements}

\bibliographystyle{aa}
\bibliography{pushkarev}

%\end{document}

\Online
%%%% TABLES %%%%
% Table 1 available electronically only

\onllongtab{1}{
\begin{longtable}{ccrrrrrrrrrr}
\caption{Derived core shift vectors.}
\label{t:core_shift}
\\
\hline\hline
  Source   &      Epoch &   median & \multicolumn{3}{c}{15.4--8.1 GHz core shift} & \multicolumn{3}{c}{15.4--8.4 GHz core shift} & \multicolumn{3}{c}{15.4--12.1 GHz core shift} \\
           &            &   jet PA &    PA  &    total &     proj &    PA  &    total &     proj &    PA  &    total &     proj  \\
           &            &    (deg) &  (deg) &    (mas) &    (mas) &  (deg) &    (mas) &    (mas) &  (deg) &    (mas) &    (mas)  \\
    (1)    &       (2)  &      (3) &    (4) &      (5) &      (6) &    (7) &      (8) &      (9) &   (10) &     (11) &     (12)  \\
\hline
\endfirsthead
\caption{Continued.} \\
\hline\hline
  Source   &      Epoch &   median & \multicolumn{3}{c}{15.4--8.1 GHz core shift} & \multicolumn{3}{c}{15.4--8.4 GHz core shift} & \multicolumn{3}{c}{15.4--12.1 GHz core shift} \\
           &            &   jet PA &    PA  &    total &     proj &    PA  &    total &     proj &    PA  &    total &     proj  \\
           &            &    (deg) &  (deg) &    (mas) &    (mas) &  (deg) &    (mas) &    (mas) &  (deg) &    (mas) &    (mas)  \\
    (1)    &       (2)  &      (3) &    (4) &      (5) &      (6) &    (7) &      (8) &      (9) &   (10) &     (11) &     (12)  \\	   
\hline
\endhead
\hline
\endfoot
\hline
\endlastfoot
0003$-$066 & 2006-07-07 &  $-$82.1 &  $-$60.3 &  0.035 &     0.033 &  $-$25.8 &  0.019 &     0.011 &  $-$62.5 &  0.015 &     0.014 \\
0003$+$380 & 2006-03-09 &    117.0 &     77.2 &  0.134 &     0.103 &     79.6 &  0.139 &     0.110 &     77.2 &  0.124 &     0.095 \\
0003$+$380 & 2006-12-01 &    116.2 &    115.5 &  0.063 &     0.063 &    121.7 &  0.106 &     0.106 &    103.0 &  0.046 &     0.044 \\
0007$+$106 & 2006-06-15 &  $-$67.6 &  $-$88.5 &  0.008 &     0.007 &  $-$20.1 &  0.011 &     0.007 &    146.2 &  0.008 &  $-$0.007 \\
0010$+$405 & 2006-04-05 &  $-$31.8 &  $-$38.3 &  0.013 &     0.013 &     64.7 &  0.008 &  $-$0.001 &   \ldots & \ldots &    \ldots \\
0010$+$405 & 2006-12-01 &  $-$32.6 &   $-$0.3 &  0.005 &     0.004 &     39.1 &  0.005 &     0.001 &  $-$89.9 &  0.010 &     0.006 \\
0055$+$300 & 2006-02-12 &  $-$50.1 &  $-$45.2 &  0.179 &     0.179 &  $-$10.2 &  0.083 &     0.064 &  $-$61.5 &  0.053 &     0.052 \\
0106$+$013 & 2006-07-07 & $-$125.2 & $-$113.7 &  0.005 &     0.005 & $-$139.2 &  0.005 &     0.005 &     20.2 &  0.002 &  $-$0.001 \\
0109$+$224 & 2006-05-24 &     85.1 &     65.3 &  0.147 &     0.138 &     91.6 &  0.073 &     0.072 &     75.8 &  0.120 &     0.118 \\
0111$+$021 & 2006-03-09 &    129.8 &    114.4 &  0.137 &     0.132 &    158.1 &  0.174 &     0.154 &    140.4 &  0.087 &     0.086 \\
0119$+$115 & 2006-06-15 &      3.7 &   $-$1.0 &  0.347 &     0.346 &   $-$0.7 &  0.300 &     0.299 &  $-$17.0 &  0.221 &     0.207 \\
0133$+$476 & 2006-08-09 &  $-$34.7 &  $-$22.4 &  0.131 &     0.128 &  $-$22.1 &  0.068 &     0.066 &  $-$93.4 &  0.025 &     0.013 \\
0149$+$218 & 2006-02-12 &  $-$13.9 &   $-$7.3 &  0.168 &     0.167 &   $-$5.4 &  0.223 &     0.221 &  $-$23.3 &  0.107 &     0.106 \\
0202$+$149 & 2006-09-06 &  $-$41.5 &      3.0 &  0.122 &     0.087 &  $-$24.4 &  0.110 &     0.105 &  $-$20.2 &  0.164 &     0.153 \\
0202$+$319 & 2006-08-09 &   $-$7.9 &    158.2 &  0.013 &  $-$0.012 &     96.9 &  0.003 &  $-$0.001 &   \ldots & \ldots &    \ldots \\
0212$+$735 & 2006-07-07 &    114.8 &    108.0 &  0.149 &     0.148 &    112.9 &  0.138 &     0.137 &     79.2 &  0.054 &     0.044 \\
0215$+$015 & 2006-04-28 &    105.7 &     89.3 &  0.088 &     0.084 &    127.0 &  0.147 &     0.137 & $-$143.6 &  0.003 &  $-$0.001 \\
0215$+$015 & 2006-12-01 &    114.5 &     76.6 &  0.241 &     0.190 &    106.4 &  0.167 &     0.165 &     95.2 &  0.059 &     0.056 \\
0219$+$428 & 2006-04-05 & $-$173.2 & $-$170.7 &  0.080 &     0.080 & $-$159.0 &  0.076 &     0.073 & $-$154.0 &  0.031 &     0.029 \\
0219$+$428 & 2006-11-10 & $-$172.9 &    179.5 &  0.173 &     0.171 &    175.4 &  0.128 &     0.125 &    179.1 &  0.179 &     0.177 \\
0224$+$671 & 2006-10-06 &      8.2 &     20.9 &  0.134 &     0.131 &     19.9 &  0.143 &     0.140 &   $-$0.9 &  0.125 &     0.124 \\
0234$+$285 & 2006-09-06 &  $-$10.1 &      1.7 &  0.275 &     0.269 &     29.6 &  0.218 &     0.168 &      1.4 &  0.171 &     0.167 \\
0235$+$164 & 2006-06-15 &  $-$15.2 &     18.7 &  0.004 &     0.003 &   $-$0.8 &  0.007 &     0.006 &   \ldots & \ldots &    \ldots \\
0241$+$622 & 2006-04-05 &    125.3 &    123.6 &  0.525 &     0.524 &    136.9 &  0.449 &     0.440 &   \ldots & \ldots &    \ldots \\
0300$+$470 & 2006-11-10 &    142.7 &    147.9 &  0.298 &     0.297 &    147.4 &  0.312 &     0.311 &    161.7 &  0.094 &     0.089 \\
0305$+$039 & 2006-02-12 &     54.8 &     53.2 &  0.303 &     0.303 &     55.3 &  0.271 &     0.271 &     60.6 &  0.127 &     0.127 \\
0309$+$411 & 2006-04-28 &  $-$56.8 &     36.6 &  0.005 &  $-$0.000 &     97.0 &  0.005 &  $-$0.004 &    118.5 &  0.007 &  $-$0.007 \\
0333$+$321 & 2006-07-07 &    125.5 &    142.9 &  0.279 &     0.266 &    143.4 &  0.273 &     0.259 &    139.1 &  0.087 &     0.084 \\
0336$-$019 & 2006-08-09 &     57.3 &     97.4 &  0.117 &     0.089 &    117.2 &  0.096 &     0.048 &     83.8 &  0.131 &     0.117 \\
0403$-$132 & 2006-05-24 &    173.3 & $-$179.6 &  0.346 &     0.343 & $-$179.6 &  0.224 &     0.222 &   \ldots & \ldots &    \ldots \\
0415$+$379 & 2006-05-24 &     65.2 &     75.8 &  0.315 &     0.310 &     79.2 &  0.191 &     0.185 &     90.3 &  0.122 &     0.111 \\
0420$-$014 & 2006-10-06 & $-$167.3 &    171.3 &  0.267 &     0.248 &    168.6 &  0.244 &     0.223 &    175.5 &  0.099 &     0.094 \\
0430$+$052 & 2006-05-24 & $-$116.3 & $-$163.6 &  0.075 &     0.051 & $-$171.5 &  0.071 &     0.041 & $-$144.2 &  0.118 &     0.104 \\
0430$+$289 & 2006-04-28 &     47.0 &     70.4 &  0.053 &     0.048 &     75.5 &  0.047 &     0.041 &     68.8 &  0.055 &     0.051 \\
0440$-$003 & 2006-07-07 & $-$134.7 &     81.2 &  0.008 &  $-$0.006 &    115.6 &  0.008 &  $-$0.003 &  $-$15.8 &  0.007 &  $-$0.004 \\
0446$+$112 & 2006-09-06 &    113.2 &    158.8 &  0.001 &     0.001 &  $-$73.6 &  0.010 &  $-$0.010 &   \ldots & \ldots &    \ldots \\
0454$+$844 & 2006-03-09 &    164.8 & $-$163.3 &  0.344 &     0.292 & $-$147.5 &  0.266 &     0.179 & $-$132.9 &  0.127 &     0.059 \\
0458$-$020 & 2006-11-10 &  $-$50.7 &    169.1 &  0.006 &  $-$0.005 & $-$171.6 &  0.004 &  $-$0.002 &  $-$61.1 &  0.000 &     0.000 \\
0528$+$134 & 2006-10-06 &     49.5 &     39.0 &  0.167 &     0.164 &     44.5 &  0.134 &     0.133 &     55.5 &  0.095 &     0.095 \\
0529$+$075 & 2006-08-09 &  $-$23.6 & $-$158.1 &  0.011 &  $-$0.008 &    142.5 &  0.001 &  $-$0.001 &    172.1 &  0.003 &  $-$0.003 \\
0552$+$398 & 2006-07-07 &  $-$74.1 &  $-$84.5 &  0.007 &     0.007 &  $-$90.4 &  0.008 &     0.007 &   \ldots & \ldots &    \ldots \\
0605$-$085 & 2006-11-10 &    123.7 &    133.1 &  0.092 &     0.091 &    156.5 &  0.103 &     0.086 &    150.7 &  0.105 &     0.093 \\
0607$-$157 & 2006-09-06 &     57.3 &     95.5 &  0.240 &     0.188 &     95.2 &  0.268 &     0.212 &     94.6 &  0.065 &     0.052 \\
0642$+$449 & 2006-10-06 &     96.1 & $-$161.4 &  0.011 &  $-$0.002 &     63.7 &  0.002 &     0.001 & $-$133.2 &  0.006 &  $-$0.004 \\
0648$-$165 & 2006-12-01 &  $-$69.5 & $-$106.7 &  0.225 &     0.179 & $-$107.4 &  0.206 &     0.163 &  $-$85.5 &  0.062 &     0.060 \\
0707$+$476 & 2006-04-05 &   $-$1.9 &  $-$17.7 &  0.153 &     0.147 &  $-$25.4 &  0.200 &     0.183 &  $-$52.2 &  0.073 &     0.046 \\
0716$+$714 & 2006-05-24 &     22.9 &  $-$16.8 &  0.127 &     0.097 &     33.8 &  0.150 &     0.148 &     15.7 &  0.198 &     0.196 \\
0723$-$008 & 2006-07-07 &  $-$44.9 &  $-$17.6 &  0.001 &     0.001 &  $-$58.9 &  0.009 &     0.009 &    127.9 &  0.002 &  $-$0.002 \\
0727$-$115 & 2006-10-06 &  $-$63.9 &  $-$63.7 &  0.240 &     0.240 &  $-$63.2 &  0.246 &     0.246 &  $-$42.0 &  0.198 &     0.183 \\
0730$+$504 & 2006-05-24 & $-$146.9 & $-$145.7 &  0.262 &     0.262 & $-$152.2 &  0.249 &     0.248 & $-$131.3 &  0.051 &     0.049 \\
0735$+$178 & 2006-04-28 &     61.8 &     79.7 &  0.039 &     0.037 &     80.0 &  0.158 &     0.150 &     74.8 &  0.157 &     0.153 \\
0736$+$017 & 2006-06-15 &  $-$73.6 &  $-$91.7 &  0.079 &     0.075 &     98.4 &  0.006 &  $-$0.006 & $-$162.8 &  0.002 &     0.000 \\
0738$+$313 & 2006-09-06 &    170.8 &    168.2 &  0.183 &     0.183 &    179.0 &  0.093 &     0.092 &    174.2 &  0.100 &     0.099 \\
0748$+$126 & 2006-08-09 &    101.1 &     89.6 &  0.098 &     0.096 &     91.0 &  0.095 &     0.094 &     91.2 &  0.041 &     0.041 \\
0754$+$100 & 2006-04-28 &     19.2 &     12.7 &  0.266 &     0.264 &     12.7 &  0.294 &     0.293 &     19.6 &  0.073 &     0.073 \\
0804$+$499 & 2006-10-06 &    106.1 &    169.0 &  0.094 &     0.043 &    165.1 &  0.052 &     0.027 &    148.6 &  0.032 &     0.024 \\
0805$-$077 & 2006-05-24 &  $-$29.4 &  $-$14.8 &  0.207 &     0.201 &  $-$41.2 &  0.260 &     0.255 &  $-$10.1 &  0.162 &     0.153 \\
0808$+$019 & 2006-08-09 &    168.7 &   \ldots & \ldots &    \ldots &   \ldots & \ldots &    \ldots &    101.7 &  0.010 &     0.004 \\
0814$+$425 & 2006-11-10 &     99.2 &     89.1 &  0.145 &     0.143 &    110.9 &  0.124 &     0.122 &     91.4 &  0.090 &     0.089 \\
0823$+$033 & 2006-06-15 &     30.2 &      9.5 &  0.141 &     0.132 &      9.2 &  0.143 &     0.133 &     16.3 &  0.083 &     0.080 \\
0827$+$243 & 2006-05-24 &    116.6 &    126.4 &  0.150 &     0.148 &    129.2 &  0.127 &     0.124 &    114.0 &  0.155 &     0.155 \\
0829$+$046 & 2006-07-07 &     66.0 &     78.4 &  0.109 &     0.106 &     75.1 &  0.153 &     0.151 &   $-$7.6 &  0.025 &     0.007 \\
0834$-$201 & 2006-03-09 & $-$115.9 &  $-$76.3 &  0.147 &     0.113 &  $-$50.3 &  0.123 &     0.051 &  $-$58.4 &  0.064 &     0.034 \\
0836$+$710 & 2006-09-06 & $-$144.7 & $-$148.5 &  0.186 &     0.185 & $-$135.8 &  0.159 &     0.157 & $-$120.7 &  0.163 &     0.149 \\
0847$-$120 & 2006-12-01 &     64.0 &     64.4 &  0.007 &     0.007 &    170.1 &  0.003 &  $-$0.001 &    158.0 &  0.005 &  $-$0.000 \\
0851$+$202 & 2006-04-28 & $-$121.2 & $-$121.9 &  0.028 &     0.028 & $-$123.5 &  0.021 &     0.021 & $-$125.3 &  0.018 &     0.018 \\
0859$-$140 & 2006-02-12 &    157.3 & $-$164.5 &  0.266 &     0.209 &    155.6 &  0.167 &     0.167 &    171.4 &  0.147 &     0.142 \\
0906$+$015 & 2006-10-06 &     43.4 &     34.4 &  0.168 &     0.166 &     26.6 &  0.239 &     0.229 &     43.2 &  0.201 &     0.201 \\
0917$+$624 & 2006-08-09 &  $-$26.2 &   $-$8.4 &  0.112 &     0.107 &   $-$6.6 &  0.111 &     0.104 &     23.2 &  0.089 &     0.058 \\
0923$+$392 & 2006-07-07 &     98.9 &    141.9 &  0.042 &     0.031 &    147.7 &  0.032 &     0.021 &     88.3 &  0.168 &     0.165 \\
0945$+$408 & 2006-06-15 &    114.2 &    125.9 &  0.083 &     0.081 &    113.0 &  0.145 &     0.145 &     75.9 &  0.027 &     0.021 \\
0953$+$254 & 2006-03-09 & $-$120.3 &    166.5 &  0.019 &     0.006 & $-$140.8 &  0.023 &     0.022 &  $-$71.5 &  0.074 &     0.049 \\
0954$+$658 & 2006-04-05 &  $-$41.8 &     16.1 &  0.005 &     0.003 &     97.7 &  0.004 &  $-$0.003 &  $-$65.1 &  0.012 &     0.011 \\
0955$+$476 & 2006-11-10 &     20.9 &   \ldots & \ldots &    \ldots &   \ldots & \ldots &    \ldots &     86.9 &  0.040 &     0.016 \\
1015$+$359 & 2006-03-09 & $-$175.7 & $-$156.6 &  0.151 &     0.143 & $-$129.8 &  0.061 &     0.042 & $-$176.2 &  0.061 &     0.061 \\
1036$+$054 & 2006-05-24 &   $-$9.2 &  $-$10.1 &  0.195 &     0.195 &   $-$5.0 &  0.170 &     0.169 &   $-$3.3 &  0.043 &     0.043 \\
1038$+$064 & 2006-10-06 &    155.2 & $-$167.2 &  0.106 &     0.084 &    177.3 &  0.188 &     0.174 &    171.0 &  0.074 &     0.072 \\
1045$-$188 & 2006-09-06 &    151.5 &    157.8 &  0.156 &     0.155 &    158.9 &  0.179 &     0.177 &    159.0 &  0.218 &     0.217 \\
1055$+$018 & 2006-11-10 &  $-$56.3 &   \ldots & \ldots &    \ldots &   \ldots & \ldots &    \ldots &  $-$97.0 &  0.074 &     0.056 \\
1101$+$384 & 2006-04-05 &  $-$24.7 &   $-$0.1 &  0.230 &     0.209 &   $-$3.4 &  0.280 &     0.260 &      1.5 &  0.147 &     0.132 \\
1127$-$145 & 2006-08-09 &     82.3 &    141.7 &  0.096 &     0.049 &    130.6 &  0.082 &     0.055 &     91.5 &  0.052 &     0.051 \\
1128$-$047 & 2006-02-12 &    161.4 &    168.1 &  0.131 &     0.130 &    173.8 &  0.307 &     0.300 &    178.0 &  0.209 &     0.200 \\
1128$-$047 & 2006-12-01 &    155.2 & $-$161.0 &  0.250 &     0.181 & $-$167.4 &  0.251 &     0.200 &   \ldots & \ldots &    \ldots \\
1148$-$001 & 2006-07-07 & $-$121.9 & $-$155.8 &  0.128 &     0.106 & $-$121.8 &  0.100 &     0.100 & $-$104.5 &  0.117 &     0.112 \\
1150$+$812 & 2006-06-15 & $-$128.7 & $-$132.8 &  0.087 &     0.087 & $-$134.5 &  0.077 &     0.076 & $-$138.0 &  0.141 &     0.139 \\
1156$+$295 & 2006-09-06 &      4.6 &      0.0 &  0.162 &     0.162 &   $-$4.1 &  0.147 &     0.145 &  $-$12.7 &  0.088 &     0.084 \\
1213$-$172 & 2006-10-06 &    117.2 &    111.0 &  0.056 &     0.056 &    110.4 &  0.132 &     0.131 &    156.0 &  0.174 &     0.136 \\
1219$+$044 & 2006-05-24 &    172.1 & $-$179.4 &  0.133 &     0.131 &    179.7 &  0.206 &     0.204 &    173.1 &  0.054 &     0.054 \\
1219$+$285 & 2006-02-12 &    109.3 &     95.5 &  0.182 &     0.177 &     63.4 &  0.068 &     0.047 &    114.9 &  0.078 &     0.078 \\
1219$+$285 & 2006-11-10 &    110.2 &    101.5 &  0.199 &     0.196 &     98.7 &  0.230 &     0.226 &    118.5 &  0.062 &     0.061 \\
1222$+$216 & 2006-04-28 &      0.6 &      9.3 &  0.180 &     0.178 &     11.0 &  0.159 &     0.156 &   $-$3.1 &  0.074 &     0.073 \\
1226$+$023 & 2006-03-09 & $-$137.3 &     48.0 &  0.020 &  $-$0.020 &  $-$97.8 &  0.022 &     0.017 & $-$149.5 &  0.118 &     0.115 \\
1243$-$072 & 2006-04-05 &  $-$91.6 & $-$167.3 &  0.012 &     0.003 & $-$119.9 &  0.018 &     0.016 & $-$105.3 &  0.015 &     0.014 \\
1253$-$055 & 2006-04-05 & $-$130.2 & $-$107.4 &  0.048 &     0.045 & $-$124.8 &  0.076 &     0.076 &  $-$94.9 &  0.098 &     0.080 \\
1253$-$055 & 2006-09-06 & $-$126.7 & $-$132.6 &  0.026 &     0.026 &  $-$85.5 &  0.059 &     0.044 &  $-$87.1 &  0.118 &     0.091 \\
1302$-$102 & 2006-03-09 &     30.5 &     47.1 &  0.220 &     0.211 &     46.8 &  0.321 &     0.308 &     64.8 &  0.064 &     0.053 \\
1308$+$326 & 2006-07-07 &  $-$44.0 &  $-$53.2 &  0.143 &     0.142 &   $-$5.8 &  0.059 &     0.047 &   $-$3.5 &  0.034 &     0.026 \\
1324$+$224 & 2006-12-01 &  $-$36.4 &     23.6 &  0.001 &     0.000 & $-$167.2 &  0.006 &  $-$0.004 & $-$104.7 &  0.007 &     0.003 \\
1331$+$170 & 2006-04-05 &     18.8 &     18.0 &  0.145 &     0.145 &     14.2 &  0.153 &     0.152 &     15.8 &  0.114 &     0.114 \\
1334$-$127 & 2006-10-06 &    149.6 & $-$174.8 &  0.237 &     0.193 & $-$177.3 &  0.311 &     0.260 & $-$175.3 &  0.167 &     0.137 \\
1345$+$125 & 2006-11-10 &    164.7 &    155.1 &  0.126 &     0.124 & $-$161.5 &  0.117 &     0.097 & $-$174.0 &  0.166 &     0.155 \\
1406$-$076 & 2006-04-05 &  $-$96.3 & $-$131.6 &  0.092 &     0.075 &  $-$36.7 &  0.120 &     0.061 & $-$124.9 &  0.101 &     0.089 \\
1413$+$135 & 2006-08-09 & $-$113.5 & $-$112.3 &  0.230 &     0.230 & $-$112.3 &  0.226 &     0.225 & $-$111.2 &  0.238 &     0.238 \\
1418$+$546 & 2006-02-12 &    131.3 &    122.2 &  0.067 &     0.066 &    121.3 &  0.103 &     0.101 &    130.5 &  0.142 &     0.142 \\
1418$+$546 & 2006-11-10 &    133.7 &    135.3 &  0.076 &     0.076 &    135.2 &  0.085 &     0.085 &    131.8 &  0.047 &     0.047 \\
1458$+$718 & 2006-09-06 &    164.1 & $-$173.1 &  0.081 &     0.075 & $-$170.4 &  0.054 &     0.049 &    118.9 &  0.031 &     0.022 \\
1502$+$106 & 2006-07-07 &    122.6 &    164.6 &  0.052 &     0.039 &    159.1 &  0.059 &     0.048 &    140.8 &  0.126 &     0.120 \\
1504$-$166 & 2006-12-01 & $-$163.7 &    174.7 &  0.148 &     0.138 &    174.2 &  0.081 &     0.075 &     79.0 &  0.019 &  $-$0.009 \\
1508$-$055 & 2006-03-09 &     81.5 &     97.3 &  0.210 &     0.202 &     99.8 &  0.149 &     0.141 &     70.4 &  0.007 &     0.007 \\
1510$-$089 & 2006-04-28 &  $-$33.2 &      4.2 &  0.122 &     0.097 &  $-$16.2 &  0.184 &     0.176 &  $-$65.2 &  0.047 &     0.040 \\
1514$+$004 & 2006-04-05 &  $-$28.1 &  $-$36.5 &  0.139 &     0.138 &   $-$8.6 &  0.178 &     0.168 &  $-$20.7 &  0.190 &     0.188 \\
1514$-$241 & 2006-04-28 &    171.7 &    179.0 &  0.188 &     0.187 &    174.9 &  0.217 &     0.217 &    177.3 &  0.135 &     0.134 \\
1532$+$016 & 2006-03-09 &    139.8 &    126.6 &  0.144 &     0.140 &    123.7 &  0.103 &     0.099 &    147.2 &  0.178 &     0.177 \\
1538$+$149 & 2006-06-15 &  $-$36.5 &  $-$81.7 &  0.032 &     0.022 &  $-$47.2 &  0.127 &     0.125 &  $-$47.7 &  0.116 &     0.114 \\
1546$+$027 & 2006-08-09 &    172.3 &     87.2 &  0.010 &     0.001 &    131.9 &  0.000 &     0.000 &  $-$79.1 &  0.006 &  $-$0.002 \\
1606$+$106 & 2006-07-07 &  $-$44.3 &  $-$23.3 &  0.057 &     0.053 &      0.1 &  0.092 &     0.066 &  $-$55.0 &  0.032 &     0.032 \\
1611$+$343 & 2006-06-15 &    157.7 & $-$161.0 &  0.057 &     0.043 & $-$173.6 &  0.063 &     0.055 &    177.9 &  0.094 &     0.088 \\
1633$+$382 & 2006-09-06 &  $-$72.8 &  $-$66.6 &  0.119 &     0.119 &  $-$68.3 &  0.158 &     0.157 &  $-$66.9 &  0.157 &     0.156 \\
1637$+$574 & 2006-05-24 & $-$156.4 & $-$167.3 &  0.117 &     0.115 & $-$162.2 &  0.089 &     0.088 &    135.0 &  0.013 &     0.005 \\
1637$+$826 & 2006-03-09 &  $-$61.5 &  $-$54.0 &  0.210 &     0.208 &  $-$63.6 &  0.148 &     0.148 &  $-$55.7 &  0.155 &     0.154 \\
1638$+$398 & 2006-08-09 &  $-$89.8 &     39.4 &  0.007 &  $-$0.005 &    103.7 &  0.011 &  $-$0.011 &   \ldots & \ldots &    \ldots \\
1641$+$399 & 2006-06-15 &  $-$90.5 &  $-$90.8 &  0.211 &     0.211 &  $-$91.5 &  0.190 &     0.190 &  $-$89.0 &  0.121 &     0.121 \\
1642$+$690 & 2006-03-09 & $-$165.6 & $-$165.2 &  0.056 &     0.055 & $-$165.6 &  0.048 &     0.048 &   \ldots & \ldots &    \ldots \\
1652$+$398 & 2006-02-12 &    147.9 &    171.5 &  0.289 &     0.265 &    171.5 &  0.269 &     0.246 &    160.4 &  0.200 &     0.196 \\
1655$+$077 & 2006-11-10 &  $-$41.4 &  $-$82.8 &  0.080 &     0.060 &  $-$84.3 &  0.042 &     0.031 &  $-$45.3 &  0.046 &     0.046 \\
1725$+$044 & 2006-03-09 &    120.5 &    135.3 &  0.100 &     0.097 &    139.7 &  0.057 &     0.054 &    125.4 &  0.062 &     0.062 \\
1726$+$455 & 2006-09-06 & $-$113.2 &  $-$87.8 &  0.009 &     0.008 &    116.1 &  0.011 &  $-$0.007 &   \ldots & \ldots &    \ldots \\
1730$-$130 & 2006-07-07 &      8.3 &     20.9 &  0.174 &     0.170 &     21.9 &  0.216 &     0.210 &     20.8 &  0.112 &     0.109 \\
1741$-$038 & 2006-12-01 & $-$130.0 &  $-$74.7 &  0.001 &     0.000 &     52.6 &  0.014 &  $-$0.014 &  $-$63.3 &  0.002 &     0.001 \\
1749$+$096 & 2006-06-15 &     25.4 &     26.9 &  0.061 &     0.061 &     11.9 &  0.106 &     0.103 &     16.3 &  0.123 &     0.121 \\
1749$+$701 & 2006-04-05 &  $-$59.4 &  $-$34.9 &  0.196 &     0.179 &  $-$64.0 &  0.224 &     0.223 &  $-$37.5 &  0.184 &     0.170 \\
1751$+$288 & 2006-10-06 &      2.2 &  $-$87.6 &  0.007 &     0.000 & $-$105.1 &  0.010 &  $-$0.003 &   \ldots & \ldots &    \ldots \\
1758$+$388 & 2006-11-10 &  $-$92.4 &  $-$31.5 &  0.079 &     0.038 &   \ldots & \ldots &    \ldots &  $-$44.9 &  0.079 &     0.053 \\
1803$+$784 & 2006-09-06 &  $-$86.0 & $-$112.7 &  0.029 &     0.026 &  $-$70.5 &  0.067 &     0.064 &  $-$91.4 &  0.076 &     0.076 \\
1807$+$698 & 2006-02-12 & $-$101.3 & $-$104.9 &  0.249 &     0.248 &  $-$92.8 &  0.186 &     0.184 &     73.3 &  0.019 &  $-$0.019 \\
1823$+$568 & 2006-07-07 & $-$160.8 &    177.1 &  0.052 &     0.048 & $-$156.6 &  0.140 &     0.140 &    101.1 &  0.013 &  $-$0.002 \\
1828$+$487 & 2006-08-09 &  $-$40.5 &  $-$53.7 &  0.117 &     0.114 &  $-$59.1 &  0.075 &     0.071 &  $-$60.7 &  0.060 &     0.056 \\
1845$+$797 & 2006-02-12 &  $-$37.3 &  $-$17.8 &  0.084 &     0.079 &  $-$62.7 &  0.111 &     0.100 &  $-$34.1 &  0.140 &     0.139 \\
1849$+$670 & 2006-05-24 &  $-$41.0 &  $-$29.7 &  0.024 &     0.024 &  $-$19.4 &  0.010 &     0.010 &  $-$24.6 &  0.014 &     0.013 \\
1901$+$319 & 2006-02-12 &    119.6 &    107.2 &  0.283 &     0.277 &    148.6 &  0.222 &     0.194 &     52.2 &  0.022 &     0.008 \\
1908$-$201 & 2006-03-09 &      6.4 &      4.5 &  0.246 &     0.246 &      1.9 &  0.219 &     0.219 &   $-$7.6 &  0.189 &     0.183 \\
1928$+$738 & 2006-04-28 &    160.7 &    163.1 &  0.147 &     0.147 &    162.9 &  0.163 &     0.162 & $-$160.3 &  0.012 &     0.009 \\
1936$-$155 & 2006-07-07 &    122.1 &    139.0 &  0.215 &     0.206 &    143.4 &  0.258 &     0.240 &    153.6 &  0.138 &     0.118 \\
1958$-$179 & 2006-10-06 &    122.9 &   $-$2.5 &  0.003 &  $-$0.002 &   \ldots & \ldots &    \ldots &   \ldots & \ldots &    \ldots \\
2005$+$403 & 2006-09-06 &    107.0 &     94.9 &  0.280 &     0.274 &     89.2 &  0.336 &     0.320 &    111.9 &  0.122 &     0.121 \\
2008$-$159 & 2006-11-10 &     21.8 &     51.9 &  0.008 &     0.007 & $-$152.6 &  0.013 &  $-$0.013 &   \ldots & \ldots &    \ldots \\
2021$+$317 & 2006-08-09 &    156.5 &    173.5 &  0.384 &     0.367 &    157.9 &  0.218 &     0.218 &    166.3 &  0.120 &     0.118 \\
2022$-$077 & 2006-04-05 &   $-$4.8 &    111.4 &  0.006 &  $-$0.003 &     78.5 &  0.007 &     0.001 &   $-$7.4 &  0.008 &     0.008 \\
2037$+$511 & 2006-05-24 & $-$142.1 &    122.5 &  0.024 &  $-$0.002 & $-$136.7 &  0.051 &     0.051 & $-$119.5 &  0.052 &     0.048 \\
2113$+$293 & 2006-02-12 & $-$170.8 & $-$165.4 &  0.268 &     0.267 & $-$164.0 &  0.225 &     0.224 & $-$169.3 &  0.114 &     0.114 \\
2121$+$053 & 2006-06-15 &  $-$95.4 & $-$127.5 &  0.152 &     0.129 & $-$107.1 &  0.149 &     0.146 &  $-$83.0 &  0.102 &     0.100 \\
2128$-$123 & 2006-10-06 & $-$149.0 & $-$142.5 &  0.223 &     0.221 & $-$143.7 &  0.261 &     0.260 & $-$125.8 &  0.055 &     0.050 \\
2131$-$021 & 2006-08-09 &     93.0 &     83.7 &  0.089 &     0.088 &     88.6 &  0.109 &     0.109 &    126.4 &  0.063 &     0.053 \\
2134$+$004 & 2006-07-07 &  $-$76.7 &  $-$83.6 &  0.188 &     0.186 &  $-$94.7 &  0.158 &     0.150 &   \ldots & \ldots &    \ldots \\
2136$+$141 & 2006-09-06 &  $-$98.3 &  $-$88.7 &  0.008 &     0.008 & $-$126.5 &  0.011 &     0.009 &   \ldots & \ldots &    \ldots \\
2145$+$067 & 2006-10-06 &    129.1 &    152.0 &  0.008 &     0.007 &     12.6 &  0.007 &  $-$0.003 &   \ldots & \ldots &    \ldots \\
2155$-$152 & 2006-12-01 & $-$139.0 & $-$126.6 &  0.405 &     0.395 & $-$149.6 &  0.296 &     0.291 & $-$143.2 &  0.286 &     0.286 \\
2200$+$420 & 2006-04-05 & $-$175.4 &    142.0 &  0.032 &     0.023 &    166.0 &  0.074 &     0.070 & $-$177.6 &  0.160 &     0.160 \\
2200$+$420 & 2006-11-10 & $-$167.4 & $-$173.6 &  0.031 &     0.031 &    178.7 &  0.135 &     0.131 & $-$171.4 &  0.124 &     0.124 \\
2201$+$171 & 2006-05-24 &     48.6 &     34.2 &  0.380 &     0.368 &     33.8 &  0.358 &     0.346 &     20.7 &  0.136 &     0.121 \\
2201$+$315 & 2006-10-06 & $-$140.4 & $-$137.7 &  0.347 &     0.346 & $-$137.7 &  0.343 &     0.343 & $-$147.5 &  0.110 &     0.109 \\
2209$+$236 & 2006-12-01 &     36.6 &  $-$71.0 &  0.038 &  $-$0.012 &    101.8 &  0.085 &     0.036 &   \ldots & \ldots &    \ldots \\
2216$-$038 & 2006-08-09 & $-$165.8 &     68.0 &  0.011 &  $-$0.006 &  $-$47.1 &  0.005 &  $-$0.002 &  $-$18.0 &  0.010 &  $-$0.008 \\
2223$-$052 & 2006-10-06 &     96.3 &     97.1 &  0.199 &     0.199 &     89.1 &  0.125 &     0.124 &    153.0 &  0.098 &     0.054 \\
2227$-$088 & 2006-07-07 &  $-$18.4 &      1.4 &  0.186 &     0.175 &      2.6 &  0.200 &     0.187 &   $-$0.8 &  0.209 &     0.199 \\
2230$+$114 & 2006-02-12 &    150.0 &    166.5 &  0.278 &     0.266 &    175.1 &  0.363 &     0.329 &    147.2 &  0.111 &     0.111 \\
2243$-$123 & 2006-09-06 &     13.2 &      6.9 &  0.161 &     0.160 &      1.6 &  0.167 &     0.163 &      7.2 &  0.140 &     0.139 \\
2251$+$158 & 2006-03-09 &  $-$83.5 & $-$104.6 &  0.124 &     0.116 & $-$111.6 &  0.115 &     0.101 &  $-$99.9 &  0.053 &     0.050 \\
2251$+$158 & 2006-06-15 &  $-$83.9 &  $-$85.0 &  0.177 &     0.177 & $-$110.7 &  0.150 &     0.134 &  $-$96.2 &  0.114 &     0.112 \\
2320$-$035 & 2006-04-05 &  $-$23.7 &  $-$53.5 &  0.009 &     0.008 &  $-$36.9 &  0.040 &     0.038 &  $-$35.8 &  0.044 &     0.043 \\
2345$-$167 & 2006-11-10 &    120.8 &    132.9 &  0.167 &     0.163 &    125.2 &  0.148 &     0.148 &    135.6 &  0.166 &     0.160 \\
2351$+$456 & 2006-05-24 &  $-$79.2 & $-$113.1 &  0.196 &     0.163 & $-$119.7 &  0.146 &     0.111 & $-$112.6 &  0.061 &     0.051 \\
2356$+$196 & 2006-04-05 & $-$149.0 & $-$156.4 &  0.203 &     0.201 & $-$154.9 &  0.185 &     0.184 & $-$174.6 &  0.166 &     0.150 \\
2356$+$196 & 2006-10-06 & $-$147.3 & $-$178.2 &  0.163 &     0.140 & $-$117.1 &  0.118 &     0.102 &    176.7 &  0.135 &     0.109 \\
\end{longtable}
{\footnotesize \noindent
Columns are as follows:
(1) IAU name (B1950.0);
(2) epoch of observations;
(3) 15.4~GHz median jet position angle;
(4) position angle of the 15.4--8.1~GHz core shift vector;
(5) magnitude of the 15.4--8.1~GHz core shift vector;
(6) 15.4--8.1~GHz core shift vector in projection on the median position angle;
(7), (8), (9), and (10), (11), (12) the same as (4), (5), (6) but for 15.4--8.1~GHz and 15.4--12.1~GHz core shifts, respectively.
}
}% End onllongtab

\onllongtab{4}{
\begin{longtable}{ccccrrrrrrc}
\caption{Derived jet parameters.}
\label{t:b_field}
\\
\hline\hline
  Source   &      Epoch & Opt. & $z$ & $\beta_\mathrm{app}$\footnote{From \cite{MOJAVE}} & $\Delta r_\mathrm{core,\,15-8\,GHz}$ & $\Omega_{r\nu}$ & $B_1$ & $B_\mathrm{core}$ & $r_\mathrm{core}$ \\
           &            &  cl. &     &                     &                                 &                 &       &                  &  \\
           &            &      &     &                     &                           (mas) &        (pc GHz) &   (G) &              (G) &             (pc) \\
\hline
\endfirsthead
\caption{Continued.} \\
\hline\hline
  Source   &      Epoch & Opt. & $z$ & $\beta_\mathrm{app}$\footnote{From \cite{MOJAVE}} & $\Delta r_\mathrm{core,\,15-8\,GHz}$ & $\Omega_{r\nu}$ & $B_1$ & $B_\mathrm{core}$ & $r_\mathrm{core}$ \\
           &            &  cl. &     &                     &                                 &                 &       &                  &  \\
           &            &      &     &                     &                           (mas) &        (pc GHz) &   (G) &              (G) &             (pc) \\
\hline
\endhead
\hline
\endfoot
\hline
\endlastfoot
0003$-$066 & 2006-07-07 & B & 0.347 &  2.89 & $<$0.051 & $<$4.55 & $<$0.20 & $>$0.22 &  $<$0.91\\
0106$+$013 & 2006-07-07 & Q & 2.099 & 26.50 & $<$0.051 & $<$7.86 & $<$0.79 & $>$0.06 & $<$13.56\\
0119$+$115 & 2006-06-15 & Q & 0.570 & 17.10 &    0.324 &   37.75 &    1.63 &    0.04 &    42.07\\
0133$+$476 & 2006-08-09 & Q & 0.859 & 12.98 &    0.099 &   13.69 &    0.77 &    0.07 &    11.60\\
0149$+$218 & 2006-02-12 & Q & 1.320 & 18.55 &    0.196 &   29.64 &    1.69 &    0.05 &    35.83\\
0202$+$149 & 2006-09-06 & Q & 0.405 &  6.41 &    0.113 &   10.87 &    0.48 &    0.10 &     4.59\\
0212$+$735 & 2006-07-07 & Q & 2.367 &  7.64 &    0.143 &   21.24 &    1.27 &    0.12 &    10.65\\
0215$+$015 & 2006-04-28 & Q & 1.715 & 34.16 &    0.111 &   17.08 &    1.41 &    0.04 &    37.98\\
0224$+$671 & 2006-10-06 & Q & 0.523 & 11.63 &    0.139 &   15.47 &    0.75 &    0.06 &    11.75\\
0234$+$285 & 2006-09-06 & Q & 1.207 & 12.26 &    0.239 &   35.82 &    1.71 &    0.06 &    28.67\\
0333$+$321 & 2006-07-07 & Q & 1.259 & 12.76 &    0.276 &   41.51 &    1.95 &    0.06 &    34.57\\
0336$-$019 & 2006-08-09 & Q & 0.852 & 22.36 &    0.105 &   14.41 &    0.92 &    0.04 &    20.99\\
0403$-$132 & 2006-05-24 & Q & 0.571 & 19.69 &    0.285 &   33.30 &    1.54 &    0.04 &    42.72\\
0420$-$014 & 2006-10-06 & Q & 0.914 &  7.36 &    0.256 &   35.94 &    1.41 &    0.08 &    17.37\\
0454$+$844 & 2006-03-09 & B & 0.112 &  0.14 &    0.302 &   10.91 &    0.27 &    0.37 &     0.72\\
0528$+$134 & 2006-10-06 & Q & 2.070 & 19.20 &    0.150 &   22.71 &    1.60 &    0.06 &    28.41\\
0552$+$398 & 2006-07-07 & Q & 2.363 &  0.36 & $<$0.051 & $<$7.72 & $<$0.36 & $>$0.68 &  $<$0.53\\
0605$-$085 & 2006-11-10 & Q & 0.872 & 19.79 &    0.096 &   13.24 &    0.84 &    0.05 &    17.07\\
0607$-$157 & 2006-09-06 & Q & 0.324 &  3.93 &    0.254 &   21.22 &    0.68 &    0.12 &     5.60\\
0716$+$714 & 2006-05-24 & B & 0.310 & 10.06 &    0.125 &   10.16 &    0.49 &    0.07 &     6.68\\
0730$+$504 & 2006-05-24 & Q & 0.720 & 14.06 &    0.255 &   33.04 &    1.47 &    0.05 &    30.30\\
0736$+$017 & 2006-06-15 & Q & 0.191 & 14.32 & $<$0.051 & $<$2.94 & $<$0.20 & $>$0.07 &  $<$2.75\\
0738$+$313 & 2006-09-06 & Q & 0.631 & 10.76 &    0.138 &   16.85 &    0.81 &    0.07 &    11.85\\
0748$+$126 & 2006-08-09 & Q & 0.889 & 18.37 &    0.097 &   13.49 &    0.84 &    0.05 &    16.15\\
0754$+$100 & 2006-04-28 & B & 0.266 & 14.40 &    0.280 &   20.38 &    0.88 &    0.05 &    19.14\\
0804$+$499 & 2006-10-06 & Q & 1.436 &  1.83 &    0.073 &   11.15 &    0.48 &    0.32 &     1.51\\
0805$-$077 & 2006-05-24 & Q & 1.837 & 50.60 &    0.228 &   34.84 &    2.71 &    0.02 &   114.73\\
0823$+$033 & 2006-06-15 & B & 0.506 & 17.80 &    0.142 &   15.57 &    0.83 &    0.05 &    18.06\\
0827$+$243 & 2006-05-24 & Q & 0.940 & 22.01 &    0.139 &   19.64 &    1.18 &    0.04 &    28.16\\
0829$+$046 & 2006-07-07 & B & 0.174 & 10.09 &    0.131 &    6.85 &    0.34 &    0.08 &     4.52\\
0836$+$710 & 2006-09-06 & Q & 2.218 & 25.38 &    0.172 &   25.72 &    1.93 &    0.05 &    42.51\\
0851$+$202 & 2006-04-28 & B & 0.306 & 15.17 & $<$0.051 & $<$4.18 & $<$0.28 & $>$0.07 &  $<$4.13\\
0859$-$140 & 2006-02-12 & Q & 1.339 & 16.47 &    0.204 &   30.96 &    1.70 &    0.05 &    33.24\\
0906$+$015 & 2006-10-06 & Q & 1.024 & 20.68 &    0.203 &   29.45 &    1.61 &    0.04 &    39.67\\
0917$+$624 & 2006-08-09 & Q & 1.446 & 15.57 &    0.111 &   16.99 &    1.09 &    0.06 &    17.25\\
0923$+$392 & 2006-07-07 & Q & 0.695 &  4.29 & $<$0.051 & $<$6.64 & $<$0.33 & $>$0.17 &  $<$1.90\\
0945$+$408 & 2006-06-15 & Q & 1.249 & 18.60 &    0.113 &   17.02 &    1.10 &    0.05 &    20.63\\
0953$+$254 & 2006-03-09 & Q & 0.712 & 11.52 & $<$0.051 & $<$6.71 & $<$0.42 & $>$0.08 &  $<$5.05\\
1015$+$359 & 2006-03-09 & Q & 1.226 & 12.46 &    0.104 &   15.55 &    0.92 &    0.07 &    12.65\\
1036$+$054 & 2006-05-24 & Q & 0.473 &  6.15 &    0.182 &   19.26 &    0.74 &    0.09 &     7.81\\
1038$+$064 & 2006-10-06 & Q & 1.265 & 11.87 &    0.146 &   21.97 &    1.19 &    0.07 &    17.03\\
1045$-$188 & 2006-09-06 & Q & 0.595 &  8.57 &    0.167 &   19.94 &    0.86 &    0.08 &    11.19\\
1101$+$384 & 2006-04-05 & B & 0.031 &  0.82 &    0.254 &    2.81 &    0.10 &    0.42 &     0.24\\
1127$-$145 & 2006-08-09 & Q & 1.184 & 14.18 &    0.089 &   13.24 &    0.84 &    0.07 &    12.25\\
1150$+$812 & 2006-06-15 & Q & 1.250 &  7.09 &    0.082 &   12.34 &    0.68 &    0.12 &     5.75\\
1156$+$295 & 2006-09-06 & Q & 0.729 & 24.73 &    0.154 &   20.11 &    1.17 &    0.04 &    32.39\\
1219$+$044 & 2006-05-24 & Q & 0.965 &  2.35 &    0.169 &   24.16 &    0.81 &    0.20 &     4.01\\
1222$+$216 & 2006-04-28 & Q & 0.432 & 21.10 &    0.170 &   17.03 &    0.90 &    0.04 &    23.41\\
1253$-$055 & 2006-09-06 & Q & 0.536 & 20.57 & $<$0.051 & $<$5.88 & $<$0.42 & $>$0.05 &  $<$7.88\\
1302$-$102 & 2006-03-09 & Q & 0.278 &  5.41 &    0.271 &   20.34 &    0.70 &    0.10 &     7.28\\
1308$+$326 & 2006-07-07 & Q & 0.997 & 27.17 &    0.095 &   13.61 &    0.96 &    0.04 &    24.08\\
1334$-$127 & 2006-10-06 & Q & 0.539 & 10.26 &    0.274 &   31.08 &    1.23 &    0.06 &    20.85\\
1413$+$135 & 2006-08-09 & B & 0.247 &  1.80 &    0.228 &   15.70 &    0.44 &    0.21 &     2.10\\
1458$+$718 & 2006-09-06 & Q & 0.904 &  7.04 &    0.068 &    9.46 &    0.51 &    0.12 &     4.38\\
1502$+$106 & 2006-07-07 & Q & 1.839 & 14.77 &    0.056 &    8.50 &    0.69 &    0.08 &     8.19\\
1504$-$166 & 2006-12-01 & Q & 0.876 &  4.31 &    0.115 &   15.90 &    0.66 &    0.15 &     4.58\\
1508$-$055 & 2006-03-09 & Q & 1.191 & 18.64 &    0.179 &   26.81 &    1.52 &    0.05 &    32.56\\
1510$-$089 & 2006-04-28 & Q & 0.360 & 20.14 &    0.151 &   13.50 &    0.73 &    0.04 &    17.71\\
1532$+$016 & 2006-03-09 & Q & 1.420 & 14.11 &    0.123 &   18.81 &    1.14 &    0.07 &    17.31\\
1538$+$149 & 2006-06-15 & B & 0.605 &  8.73 &    0.077 &    9.25 &    0.49 &    0.09 &     5.29\\
1546$+$027 & 2006-08-09 & Q & 0.414 & 12.08 & $<$0.051 & $<$5.09 & $<$0.32 & $>$0.08 &  $<$4.01\\
1606$+$106 & 2006-07-07 & Q & 1.226 & 18.91 &    0.073 &   10.97 &    0.79 &    0.06 &    13.52\\
1611$+$343 & 2006-06-15 & Q & 1.397 & 14.11 &    0.059 &    9.07 &    0.66 &    0.08 &     8.35\\
1633$+$382 & 2006-09-06 & Q & 1.814 & 29.45 &    0.139 &   21.21 &    1.62 &    0.04 &    40.67\\
1637$+$574 & 2006-05-24 & Q & 0.751 & 10.61 &    0.103 &   13.51 &    0.71 &    0.08 &     9.37\\
1641$+$399 & 2006-06-15 & Q & 0.593 & 19.27 &    0.201 &   23.85 &    1.20 &    0.04 &    29.94\\
1642$+$690 & 2006-03-09 & Q & 0.751 & 16.65 & $<$0.051 & $<$6.85 & $<$0.48 & $>$0.06 &  $<$7.43\\
1652$+$398 & 2006-02-12 & B & 0.033 &  0.21 &    0.279 &    3.25 &    0.10 &    0.48 &     0.22\\
1655$+$077 & 2006-11-10 & Q & 0.621 & 14.45 &    0.061 &    7.45 &    0.47 &    0.07 &     7.02\\
1726$+$455 & 2006-09-06 & Q & 0.717 &  1.82 & $<$0.051 & $<$6.73 & $<$0.28 & $>$0.30 &  $<$0.91\\
1730$-$130 & 2006-07-07 & Q & 0.902 & 35.69 &    0.195 &   27.31 &    1.69 &    0.03 &    63.44\\
1749$+$096 & 2006-06-15 & B & 0.322 &  6.84 &    0.083 &    6.92 &    0.33 &    0.11 &     3.11\\
1749$+$701 & 2006-04-05 & B & 0.770 &  6.03 &    0.203 &   27.03 &    1.04 &    0.10 &    10.75\\
1803$+$784 & 2006-09-06 & B & 0.680 &  8.97 & $<$0.051 & $<$6.58 & $<$0.39 & $>$0.10 &  $<$3.86\\
1807$+$698 & 2006-02-12 & B & 0.051 &  0.10 &    0.216 &    3.81 &    0.12 &    0.47 &     0.25\\
1823$+$568 & 2006-07-07 & B & 0.664 & 20.85 &    0.094 &   11.79 &    0.74 &    0.05 &    16.01\\
1828$+$487 & 2006-08-09 & Q & 0.692 & 13.65 &    0.096 &   12.24 &    0.69 &    0.06 &    10.90\\
1849$+$670 & 2006-05-24 & Q & 0.657 & 30.63 & $<$0.051 & $<$6.48 & $<$0.52 & $>$0.04 & $<$12.92\\
1901$+$319 & 2006-02-12 & Q & 0.635 &  2.67 &    0.236 &   29.03 &    0.87 &    0.16 &     5.39\\
1928$+$738 & 2006-04-28 & Q & 0.302 &  8.43 &    0.155 &   12.31 &    0.54 &    0.08 &     6.80\\
1936$-$155 & 2006-07-07 & Q & 1.657 &  2.60 &    0.236 &   36.23 &    1.31 &    0.20 &     6.57\\
2005$+$403 & 2006-09-06 & Q & 1.736 & 12.21 &    0.308 &   47.18 &    2.34 &    0.06 &    37.61\\
2037$+$511 & 2006-05-24 & Q & 1.686 &  3.30 & $<$0.051 & $<$7.98 & $<$0.45 & $>$0.25 &  $<$1.79\\
2113$+$293 & 2006-02-12 & Q & 1.514 &  1.40 &    0.247 &   37.78 &    1.16 &    0.27 &     4.23\\
2121$+$053 & 2006-06-15 & Q & 1.941 & 13.29 &    0.148 &   22.61 &    1.43 &    0.07 &    19.61\\
2128$-$123 & 2006-10-06 & Q & 0.501 &  6.94 &    0.242 &   26.41 &    0.98 &    0.08 &    12.05\\
2131$-$021 & 2006-08-09 & B & 1.285 & 20.02 &    0.099 &   14.94 &    1.02 &    0.05 &    19.49\\
2134$+$004 & 2006-07-07 & Q & 1.932 &  5.94 &    0.172 &   26.23 &    1.31 &    0.13 &    10.28\\
2136$+$141 & 2006-09-06 & Q & 2.427 &  5.43 & $<$0.051 & $<$7.68 & $<$0.55 & $>$0.20 &  $<$2.76\\
2145$+$067 & 2006-10-06 & Q & 0.990 &  2.52 & $<$0.051 & $<$7.48 & $<$0.34 & $>$0.26 &  $<$1.32\\
2155$-$152 & 2006-12-01 & Q & 0.672 & 18.11 &    0.343 &   43.23 &    1.89 &    0.04 &    51.02\\
2200$+$420 & 2006-04-05 & B & 0.069 & 10.57 &    0.052 &    1.21 &    0.09 &    0.11 &     0.84\\
2201$+$171 & 2006-05-24 & Q & 1.076 &  2.55 &    0.369 &   54.08 &    1.55 &    0.16 &     9.64\\
2201$+$315 & 2006-10-06 & Q & 0.295 &  7.87 &    0.345 &   27.04 &    0.95 &    0.07 &    13.96\\
2209$+$236 & 2006-12-01 & Q & 1.125 &  3.43 & $<$0.051 & $<$7.69 & $<$0.39 & $>$0.22 &  $<$1.79\\
2223$-$052 & 2006-10-06 & Q & 1.404 & 17.34 &    0.162 &   24.63 &    1.47 &    0.05 &    27.83\\
2227$-$088 & 2006-07-07 & Q & 1.560 &  8.14 &    0.193 &   29.60 &    1.44 &    0.09 &    15.80\\
2230$+$114 & 2006-02-12 & Q & 1.037 & 15.41 &    0.320 &   46.48 &    2.12 &    0.05 &    46.70\\
2243$-$123 & 2006-09-06 & Q & 0.632 &  5.49 &    0.164 &   20.10 &    0.78 &    0.11 &     7.30\\
2251$+$158 & 2006-06-15 & Q & 0.859 & 14.19 &    0.159 &   22.00 &    1.13 &    0.06 &    20.36\\
2345$-$167 & 2006-11-10 & Q & 0.576 & 13.45 &    0.157 &   18.44 &    0.90 &    0.06 &    16.18\\
2351$+$456 & 2006-05-24 & Q & 1.986 & 18.01 &    0.171 &   25.97 &    1.72 &    0.06 &    30.48\\
\end{longtable}
}% End onllongtab
\end{document}